\documentclass[lettersize,journal]{IEEEtran}
\usepackage{amsmath,amsfonts}
\usepackage{algorithmic}
\usepackage{algorithm}
\usepackage{array}
\usepackage[caption=false,font=normalsize,labelfont=sf,textfont=sf]{subfig}
\usepackage{textcomp}
\usepackage{stfloats}
\usepackage{url}
\usepackage{verbatim}
\usepackage{graphicx}
\usepackage{cite}

\usepackage{tablefootnote}
\usepackage{multirow}
\usepackage[table, dvipsnames]{xcolor}
\usepackage{booktabs} 
\usepackage{wasysym}

\begin{document}





\newpage
\title{\textit{Golden Gemini} is All You Need: Finding the Sweet Spots for Speaker Verification}

\author{Tianchi Liu,~\IEEEmembership{Student Member,~IEEE}, Kong Aik Lee,~\IEEEmembership{Senior Member,~IEEE}, Qiongqiong Wang,~\IEEEmembership{Member,~IEEE}, Haizhou Li,~\IEEEmembership{Fellow,~IEEE}

\thanks{Manuscript received 17 October 2023; revised 14 February 2024; accepted 17 March 2024. Date of publication 12 April 2024; date of current version 19 April 2024.
This work was supported in part by the Agency for Science, Technology and Research (A$^\star$STAR), Singapore, through its Council Research Fund (Project No. CR-2021-005), in part by the National Natural Science Foundation of China (Grant No. 62271432), in part by the Shenzhen Science and Technology Research Fund (Fundamental Research Key Project Grant No. JCYJ20220818103001002), and in part by the Internal Project Fund from Shenzhen Research Institute of Big Data under Grant No. T00120220002.
(Corresponding Author: Kong Aik Lee)
}

\thanks{Tianchi Liu and Qiongqiong Wang are with Institute for Infocomm Research (I$^2$R), Agency for Science, Technology and Research (A$^\star$STAR), Singapore. (e-mail: \{liu\underline{\enskip}tianchi, wang\underline{\enskip}qiongqiong\}@i2r.a-star.edu.sg)}

\thanks{Tianchi Liu is also with the Department
 of Electrical and Computer Engineering, National University of Singapore, Singapore.}
\thanks{Kong Aik Lee is with the Department of Electrical and Electronic Engineering, The Hong Kong Polytechnic University, Hong Kong. (e-mail: kong-aik.lee@polyu.edu.hk).}
\thanks{Haizhou Li is with 
Shenzhen Research Institute of Big Data, School of Data Science, The Chinese University of Hong Kong, Shenzhen, China, and the Department of Electrical and Computer Engineering, National University of Singapore, Singapore. (e-mail: haizhouli@cuhk.edu.cn).}

}

\markboth{Accepted for publication in IEEE/ACM TRANSACTIONS ON AUDIO, SPEECH, AND LANGUAGE PROCESSING}%
{Shell \MakeLowercase{\textit{et al.}}: A Sample Article Using IEEEtran.cls for IEEE Journals}

\IEEEpubid{$\copyright$2024 The Authors. This work is licensed under a Creative Commons Attribution-NonCommercial-NoDerivatives 4.0 License. }

\maketitle

\begin{abstract} 
The residual neural networks (ResNet) demonstrate the impressive performance in automatic speaker verification (ASV). They treat the time and frequency dimensions equally, following the default stride configuration designed for image recognition, where the horizontal and vertical axes exhibit similarities.
This approach ignores the fact that time and frequency are asymmetric in speech representation.
{We address this issue and postulate \textit{Golden-Gemini Hypothesis,} which posits the prioritization of temporal resolution over frequency resolution for ASV. 
The hypothesis is verified by conducting a systematic study on the impact of temporal and frequency resolutions on the performance}, using a trellis diagram to represent the stride space. We further identify two optimal points, namely \textit{Golden Gemini}, which serves as a guiding principle for designing 2D ResNet-based ASV models.
By following the principle, a state-of-the-art ResNet baseline model gains a significant performance improvement on VoxCeleb, SITW, and CNCeleb datasets with 7.70\%/11.76\% average EER/minDCF reductions, respectively, across different network depths (ResNet18, 34, 50, and 101), while reducing the number of parameters by 16.5\% and FLOPs by 4.1\%. We refer to it as \textit{Gemini} ResNet.
Further investigation reveals the efficacy 
of the proposed \textit{Golden Gemini} operating points across various training conditions and architectures. 
Furthermore, we present a new benchmark, namely the \textit{Gemini} DF-ResNet, using a cutting-edge model. 
\textit{Codes and pre-trained models are available at https://github.com/Tianchi-Liu9/Golden-Gemini-for-Speaker-Verification.}
\end{abstract}

\begin{IEEEkeywords}
Speaker verification, speaker recognition, 2D CNN, ResNet, stride configuration, temporal resolution.
\end{IEEEkeywords}

\section{Introduction}
\label{sec_Introduction}
\IEEEPARstart{A}{utomatic} speaker verification (ASV) aims to verify the claimed identity of a speaker according to his/her voice~\cite{RSR2015_dataset}. Currently, deep learning-based speaker embedding has emerged as the predominant method~\cite{xvector}.
In this approach, fixed-dimensional representations are extracted from enrollment and test speech utterances~\cite{7846260Snyder}. These representations, rich in voice characteristics, are referred to as speaker embeddings~\cite{xvector}. The neural networks responsible for extracting these embeddings are known as the embedding extractors. The recognition procedure is often done by measuring the similarity between embeddings, using methods such as cosine similarity or probabilistic linear discriminant analysis (PLDA)~\cite{4409052, 5947437, wang2023generalized, QQ2022interspeech, 10022331, 10097032}.

Typical speaker-embedding neural networks consist of three components ~\cite{TDNN_peddinti15b_interspeech}. First, an encoder is used to extract frame-level features from an input utterance. It is followed by a temporal aggregation layer that combines the frame-level features from the encoder into a fixed-length condensed representation of the entire input sequence. Commonly used temporal aggregation techniques include average pooling~\cite{9362097}, statistical pooling ~\cite{temporal_statistics_pooling}, attentive pooling ~\cite{Okabe2018, Zhu2018}, and posterior inference~\cite{xivector}. The output stage of the neural network constitutes a decoder that classifies utterance-level representations into speaker classes~\cite{arcface, 9739948, indefence}. It utilizes a stack of fully-connected layers, including a bottleneck layer specifically designed for extracting speaker embeddings. 
Among these, the encoder is often the heaviest part of the model. The efficacy and efficiency of its design are instrumental to the performance of the model.

\IEEEpubidadjcol Many prior studies have investigated and designed numerous powerful networks as encoders. These backbone networks can be broadly categorized into four main types:
\begin{itemize}
\item 2D convolutional neural network (CNN)~\cite{BUT2019system, thin_resnet,wespeaker, MIAO2021201, indefence, resnext_n_res2net_ASV, depth_first, depth_first_conf, 10015805, chen2023enhanced, 9413948, 9746294, 9747384, 9746688}, 
\item Time-delay neural network (TDNN)~\cite{ECAPA_TDNN, 8683760,9688031, 10096814, 9306293, 9746688}, \item Transformer~\cite{9746050, Safari2020}, and
\item  Combinations of the aforementioned three~\cite{ECAPA_CNN_TDNN, MFA_TDNN, wang2023cam++, 10023305, 10096659, 10095051, PAM, zhang22h_interspeech, pvector2023intspch}. 
\end{itemize}
Among these architectures, 2D CNN is the most widely used for ASV.
It is worth mentioning that in the VoxCeleb Speaker Recognition Challenge (VoxSRC) 2021~\cite{voxsrc2021} and 2022~\cite{voxsrc2022}, the best-performing models are based on 2D CNNs, with ResNet~\cite{resnet} being the preferred choice~\cite{makarov2022id, zhao2023hccl,cai2021dku, zhao2021speakin}. 
ResNet is not only popular in ASV but also widely employed in other speech-related tasks, such as speaker extraction~\cite{SpEx+, 9721129, 10097239, jiang23c_interspeech, 10097232}, target-speaker voice activity detection\cite{9747019, 10094752, 10.1145/3474085.3475587}, speaker diarization~\cite{9413832, 10096449}, and speech anti-spoofing~\cite{9746820, 10159518, 10057965, 9413501}. 
Therefore, investigating the ResNet architecture for speech-related tasks holds significant importance.

The ResNet architecture was initially designed for image recognition~\cite{resnet} where the horizontal and vertical dimensions of images have similar implications~\cite{perez2017effectiveness, liu2022convnet} and are often uniform in size, typically $N \times N$ pixels with commonly used values such as 224 and 384. Consequently, it is intuitive to treat these two dimensions equally with the default equal-stride configuration in ResNet~\cite{resnet}. 
However, when dealing with speech representations, the time and frequency axes of speech spectrograms possess distinct implications~\cite{6853584} and often vary in size (e.g., 80 $\times$ 301~\cite{MFA_TDNN}). Therefore, the techniques that work for image recognition may not be suitable for ASV, thus necessitating appropriate modifications.
Despite these notable differences in feature properties between images and speech signals, existing ASV systems based on the ResNet models~\cite{BUT2019system, thin_resnet, 9746688, indefence, resnext_n_res2net_ASV, wespeaker, makarov2022id, zhao2023hccl,cai2021dku,zhao2021speakin, MIAO2021201,9413948, 9747384, depth_first, depth_first_conf, 10015805, chen2023enhanced, 9746294} continue to treat the frequency and temporal resolutions equally by adopting the default stride configuration as the original ResNet. Doubts arise regarding the adequacy of this equal-stride configuration for ASV.

The preservation of temporal resolution in various existing ASV methods~\cite{liu2023disentangling, ECAPA_CNN_TDNN, ECAPA_TDNN, 8683760,9688031, 10096814, 9306293, 9746688, 10023305, MFA_TDNN, wang2023cam++, Safari2020, 9746050, SUDA, SUV, pvector2023intspch} has led to the hypothesis that ASV may be more sensitive to temporal resolution than frequency resolution.
TDNN-based models~\cite{ECAPA_CNN_TDNN,ECAPA_TDNN, 8683760,9688031, 10096814, 9306293, 9746688,MFA_TDNN, wang2023cam++, 10023305, pvector2023intspch} preserve the temporal resolution across the stacked layers.
Similarly, recurrent networks, such as the long short-term memory (LSTM), preserve the number of frames~\cite{SUDA, SUV}. Recent studies ~\cite{Safari2020, 9746050, pvector2023intspch} adopt the Transformer architecture as the encoder, ensuring the preservation of the temporal resolution across stacked Transformer blocks.
Should temporal resolution prove to be of greater significance, the equal-stride configuration may not be optimal since it diminishes the temporal resolution.
{The current understanding of the impact of temporal and frequency resolutions on the performance of ResNet-based ASV models remains limited, leaving a research gap to be filled.} Consequently, this motivates us to  
explore the relative importance of temporal and frequency resolution in the feature representation process of ASV.
Building upon this investigation, we identify the optimal stride configurations that account for the inherent characteristics of speech signals to better align with the requirements of ASV, leading to improved performance. We also conduct a meticulous analysis of the trade-offs between performance and model complexity to ensure both efficacy and efficiency.
The major contributions of this work are summarized as follows:
\begin{itemize}
\item {We postulate \textbf{\textit{Golden-Gemini Hypothesis}}, which posits that the preservation of temporal resolution is to be prioritized over frequency resolution for the optimal extraction of speaker characteristics.}
\item  We systematically analyze the joint effects of temporal and frequency resolutions through a carefully designed trellis diagram. Two optimal spots on the trellis diagram are identified and named \textit{Golden Gemini}.
\item Based on the insights gained from the trellis diagram analysis, we summarize a set of guiding principles for designing ResNet-based models for ASV.
\item The compatibility and efficacy of the proposed \textit{Golden Gemini} models are evaluated under various aspects, including model sizes, structures (backbones, attention, pooling layers and micro design), training strategies, and in/cross-domain test sets. 
\item We introduce the \textit{Gemini} DF-ResNet, 
as the new state-of-the-art (SOTA) benchmark for ASV.
\end{itemize}

\section{Background}

\begin{table}[t]
\caption{A comparison between the original ResNet34~\cite{resnet}, modified ResNet~\cite{BUT2019system} and the proposed \textit{Gemini} ResNet34. A 2D CNN layer is represented in the format of [kernel size$\times$kernel size, number of channels (\textit{C})]. In the original ResNet34, \textit{C} is set to 64, while the modified ResNet and \textit{Gemini} ResNet use a value of 32. The symbol `-' indicates the layer is not employed in the model. When applicable, a (2,2) stride is performed in the first CNN layer of the stage.}
\label{table_structure}
\begin{center}
\begin{scriptsize}
\setlength{\tabcolsep}{0.5mm}{
\renewcommand{\arraystretch}{1.2}{
\begin{tabular}{c|c|cc|cc|cc}
\hline
\toprule
\multirow{2}{*}{Stage} & \multirow{2}{*}{Layer} & \multicolumn{2}{c|}{original ResNet34} & \multicolumn{2}{c|}{modified ResNet34} & \multicolumn{2}{c}{\textit{Gemini} ResNet34} \\
& & Stride & Output  & Stride & Output   & Stride & Output\\ 

\hline
\midrule

\multirow{2}{*}{\textit{conv1}} & 7$\times$7, \textit{C} & (2,2) & F/2$\times$T/2 &- & -&- & -\\
& 3$\times$3, \textit{C} & - & -  & (1,1) & F$\times$T & (1,1) & F$\times$T \\ \hline
\multirow{3}{*}{\textit{conv2}}  &  Max Pooling & (2,2) & F/4$\times$T/4  &- & -&- & - \\ 
&  $\begin{bmatrix}3\times3,  C\\3\times3, C \end{bmatrix}$$\times$3&(1,1)&F/4$\times$T/4&(1,1)&F$\times$T &(2,1)&F/2$\times$T \\ \hline
\textit{conv3}&  $\begin{bmatrix}3\times3,  C\times2\\3\times3, C\times2 \end{bmatrix}$$\times$4&(2,2)&F/8$\times$T/8&(2,2)&F/2$\times$T/2&(2,2)&F/4$\times$T/2 \\ \hline
\textit{conv4}&  $\begin{bmatrix}3\times3, C\times4\\3\times3, C\times4\end{bmatrix}$$\times$6&(2,2)&F/16$\times$T/16&(2,2)&F/4$\times$T/4&(2,1)&F/8$\times$T/2 \\ \hline
\textit{conv5}&  $\begin{bmatrix}3\times3, C\times8\\3\times3, C\times8\end{bmatrix}$$\times$3&(2,2)&F/32$\times$T/32&(2,2)&F/8$\times$T/8&(2,1)&F/16$\times$T/2 \\ 

\bottomrule
\hline

\end{tabular}
}
}
\end{scriptsize}
\end{center}
\vspace{-3 mm}
\end{table}

\begin{figure*}[t]
\centerline{\includegraphics[scale=0.065]{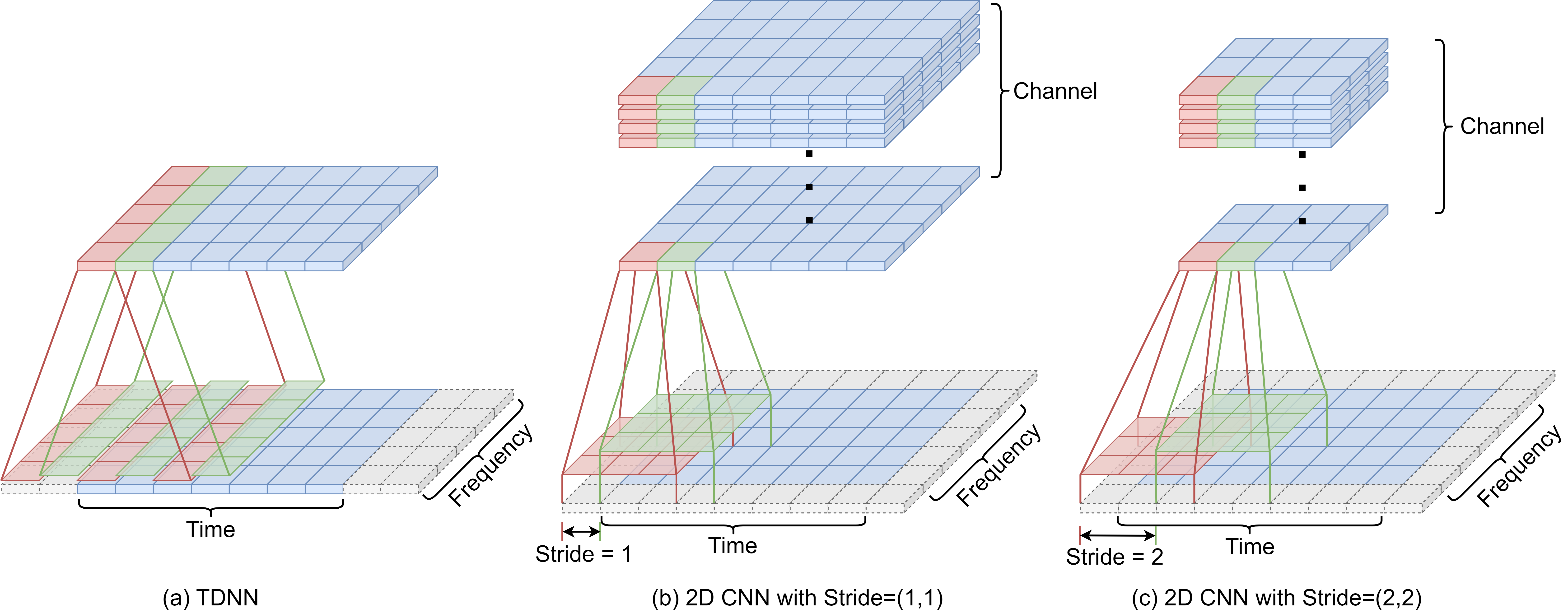}}

\vspace{-1 mm}
\caption{The illustration of convolution operations in (a) TDNN, (b) 2D CNN with stride = (1,1), and (c) stride = (2,2). The blue and grey cuboids represent time-frequency bins of feature maps and paddings, respectively.}
\label{fig_TDNN_CNN}
\vspace{-2 mm}
\end{figure*}

\subsection{ResNet Architecture}
\label{subsec_resnet}
ResNet is first proposed for image recognition~\cite{resnet}. A standard ResNet comprises five stages. The first stage is a $7 \times 7$ convolutional (\textit{conv}) layer, followed by four stages. Each stage contains multiple residual blocks, as shown in Table~\ref{table_structure}. 
Residual blocks are defined as:
\begin{equation}
\label{eq_residual_block}
    \mathrm{\textbf{y}} = \mathcal{F} (\mathrm{\textbf{x}},\{\mathrm{\textbf{W}}_{i}\}) + \mathrm{\textbf{W}}_{s}\mathrm{\textbf{x}},
\end{equation}
where $\mathrm{\textbf{x}}$ and $\mathrm{\textbf{y}}$ are the input and output vectors of a residual block. The function $\mathcal{F} (\mathrm{\textbf{x}},\{\mathrm{\textbf{W}}_{i}\})$ represents the residual mapping to be learned. The operation $\mathcal{F} + \mathrm{\textbf{x}}$
is performed by a shortcut connection and element-wise addition. 
$\mathrm{\textbf{W}}_{s}$ denotes a linear projection  
used in the shortcut to match the dimensions of $\mathrm{\textbf{x}}$ to $\mathcal{F}$. 
The design of $\mathcal{F}$ is flexible and commonly categorized into two types: basic block and bottleneck block. Basic blocks utilize two $(3 \times 3)$ convolutional layers, whereas the bottleneck blocks are composed of $(1 \times 1)$, $(3 \times 3)$, and $(1 \times 1)$ convolutions. 
The weights of these layers are denoted as $\{\mathrm{\textbf{W}}_{i}\}$, and the bias is omitted for simplicity~\cite{resnet}.

The depth of ResNet is determined by the number of layers $M$, which is dictated by the types and number $m$ of residual blocks. It is formulated as:

\begin{equation}
\label{eq_depth}
   M =  \begin{cases}2 \times m + k, & \text{if basic block} \\3 \times m + k, & \text{if bottleneck block}\end{cases},
\end{equation}
where $k$ accounts for the convolution layer in the first \textit{conv1} stage and the bottleneck layer in the decoder, typically assigned a value of 2.
ResNets with $M=18/34/50/101/152$ layers are commonly adopted~\cite{resnet}. The depth can be further extended, such as 233~\cite{depth_first} and 1202~\cite{resnet} layers. In addition to expanding the depth, previous studies explore various variations of ResNet architecture from different perspectives to improve the performance, including ResNeXt~\cite{ResNeXt}, ConvNeXt~\cite{liu2022convnet}, Res2Net~\cite{res2net}, squeeze-and-excitation network (SENet)~\cite{SEnet}, depth-first ResNet (DF-ResNet)~\cite{depth_first,depth_first_conf}, separate downsampling  ResNet (SD-ResNet)~\cite{liu2022convnet}, modified ResNet~\cite{BUT2019system, wespeaker}, thin-ResNet~\cite{thin_resnet} and fast ResNet~\cite{indefence}.

We observe that the five-stage structure remains intact, despite the adjustments to network depth or modifications to the model architecture~\cite{ResNeXt, liu2022convnet, res2net, SEnet, depth_first,depth_first_conf, BUT2019system, wespeaker, thin_resnet, indefence}. Therefore, in this work, we validate our hypothesis by adopting the five-stage design, while acknowledging that the hypothesis itself is applicable to architectures with arbitrary stages.
The generality of our proposed method allows its application to all 2D CNN models following the five-stage design, including Res2Net~\cite{res2net}, SENet~\cite{SEnet}, DF-ResNet~\cite{depth_first,depth_first_conf}, SD-ResNet~\cite{liu2022convnet}, and modified ResNet~\cite{BUT2019system, wespeaker}, as validated through experiments.

\subsection{Extensions of ResNet}
\label{Sec_mod_resnets}
The ResNet initially designed for an image recognition task~\cite{resnet}, exhibits inferior performance when directly applied to speaker verification~\cite{thin_resnet}. 
Our initial findings also suggest the same, highlighting the inherent differences between image and speech, and the necessity of customizing ResNet for speech-related tasks.

\textbf{Preserve Resolutions}. By simply removing the stride operations (2,2) in the first and second stages, a modified ResNet gains a remarkable improvement~\cite{BUT2019system, wespeaker}. A comparison of the original ResNet~\cite{resnet} and the modified structure~\cite{BUT2019system} is shown in Table~\ref{table_structure}.
We believe that removing the stride operations in the first two stages preserves the time and frequency resolutions, allowing for the extraction of low-level features. This assumption emphasizes the significance of resolutions as an important aspect of ASV. Nevertheless, it remains uncertain whether the time resolution, frequency resolution, or both are significant to the overall performance, which warrants further investigation.

\textbf{Prioritize depth over width}. Previous studies adopt a computationally efficient operation by reducing the width of ResNet~\cite{BUT2019system, 9413948, indefence}. Recent work further investigates the trade-off between the depth and width of networks, highlighting that depth plays a more important role in ASV~\cite{depth_first}. 
In this paper, we examine ResNet-based networks from a different perspective, focusing on investigating how time and frequency resolutions affect performance, as well as considering the model size and FLOPs. Our findings complement the depth-first rule~\cite{depth_first} presented in Section~\ref{subsec_new_SOTA}.

\subsection{Stride and Resolution}
\label{Sec_stride_res_TDNN_resnet}
In this subsection, we provide an overview of how the stride configuration influences the temporal and frequency resolutions in the 1D TDNN and 2D CNN models. This forms the basis for our subsequent exploration and investigation in the following sections.

As illustrated in Fig.~\ref{fig_TDNN_CNN} (a), a TDNN network implemented with dilated 1D CNN layers~\cite{ECAPA_TDNN} treats the input as 1D features, while considering the frequency dimension as channels. {TDNN-based models are not included in this study due to the absence of the frequency dimension, and existing TDNN models generally maintain time resolution}~\cite{ECAPA_CNN_TDNN,ECAPA_TDNN, 8683760,9688031, 10096814, 9306293, 9746688,MFA_TDNN, wang2023cam++, 10023305, pvector2023intspch}.
Unlike TDNNs, a 2D CNN considers the input feature as a 3-dimensional tensor $C \times F \times T$, where \textit{C}, \textit{F}, and \textit{T} represent the channel, frequency, and time dimensions, respectively~\cite{depth_first}. By employing multiple 2D CNN layers, the number of channels increases, while the frequency and temporal resolutions decrease by downsampling operations to reduce computational complexity~\cite{resnet}.
The output dimension of the downsampling operation is mainly controlled by the stride. 
Figure~\ref{fig_TDNN_CNN} (b) and (c) illustrate that by adjusting the stride in each dimension, the temporal and frequency resolutions can be controlled independently. 
For instance, setting the stride to 2 on the time dimension and 1 for the frequency dimension roughly halves the time resolution while keeping the frequency resolution the same.

In addition to stride ($S$), the output resolution ($R_{\mathrm{out}}$) is also affected by the input resolution ($R_{\mathrm{in}}$), padding ($P$), dilation ($D$), and the kernel size ($K$), as follows:

\begin{equation}
\label{eq_shape_conv}
    R_{\mathrm{out}} = \frac{R_{\mathrm{in}} + 2 \times P - D \times (K -1) -1 }{S} + 1 \simeq \frac{R_{\mathrm{in}}}{S}.
\end{equation}
In summary, the temporal and frequency resolutions are primarily controlled by the stride configuration employed on each dimension. In this paper, we investigate the impact of time and frequency resolutions on ASV performance by comparing different stride configurations. We aim to identify the optimal stride configurations for ASV.

\section{Golden-Gemini is All You Need}
\subsection{Golden-Gemini Hypothesis}
\label{sec_trellis_diagram}

Considering the distinct physical implications of the two dimensions in speech representations, we raise doubts regarding the appropriateness of employing the default equal-stride configuration, originally designed for image recognition. Furthermore, given that existing studies show the benefit of preserving the temporal resolution during the feature extraction stage~\cite{ECAPA_CNN_TDNN, ECAPA_TDNN, 8683760,9688031, 10096814, 9306293, 9746688, 10023305, MFA_TDNN, wang2023cam++, 9746050, SUDA, SUV}, we postulate the following hypothesis:

\textbf{\textit{Golden-Gemini Hypothesis: }} \textit{In the context of a ResNet architecture, characterized by a sequence of multiple stages (typically 5), there exist operational states that yield optimal performance. These states can be determined by following a temporal-frequency stride configuration that prioritizes the preservation of temporal resolution over frequency resolution.} We refer to these specific operational states as the \textit{Golden-Gemini} configurations.

The \textbf{\textit{Golden-Gemini Hypothesis}} posits that the preservation of temporal resolution is to be prioritized over frequency resolution for the optimal extraction of speaker characteristics. 

The uniqueness of a person's voice results from the combination of physiological characteristics inherent in the vocal tract and the learned speaking habits of different individuals~\cite{Speaker_Recognition_Tutorial}. 
The vocal tract shape is an important physical distinguishing factor~\cite{Speaker_Recognition_Tutorial}, wherein the laryngeal features encompass pitch and glottal pulse shape, while the supra-laryngeal features are associated with the formant frequencies, bandwidths, and intensities~\cite{1318526}. 
These features appear across various time scales, underscoring the significance of maintaining adequate temporal resolution for the convolution filters. By progressively covering larger local regions as the network deepens, these filters extract meaningful representations from neighboring frames.
On the other hand, the learned speaking habits, including speaking rate and prosodic effects~\cite{Speaker_Recognition_Tutorial}, vary along the time dimension. By preserving temporal resolution, models can effectively capture these time-dependent patterns.
Conversely, downsampling in the time domain leads to a loss of neighboring frame information and diminishes its ability to capture fine-grained details necessary for robust speaker discrimination.

\begin{figure}[t]
\centerline{\includegraphics[scale=0.062]{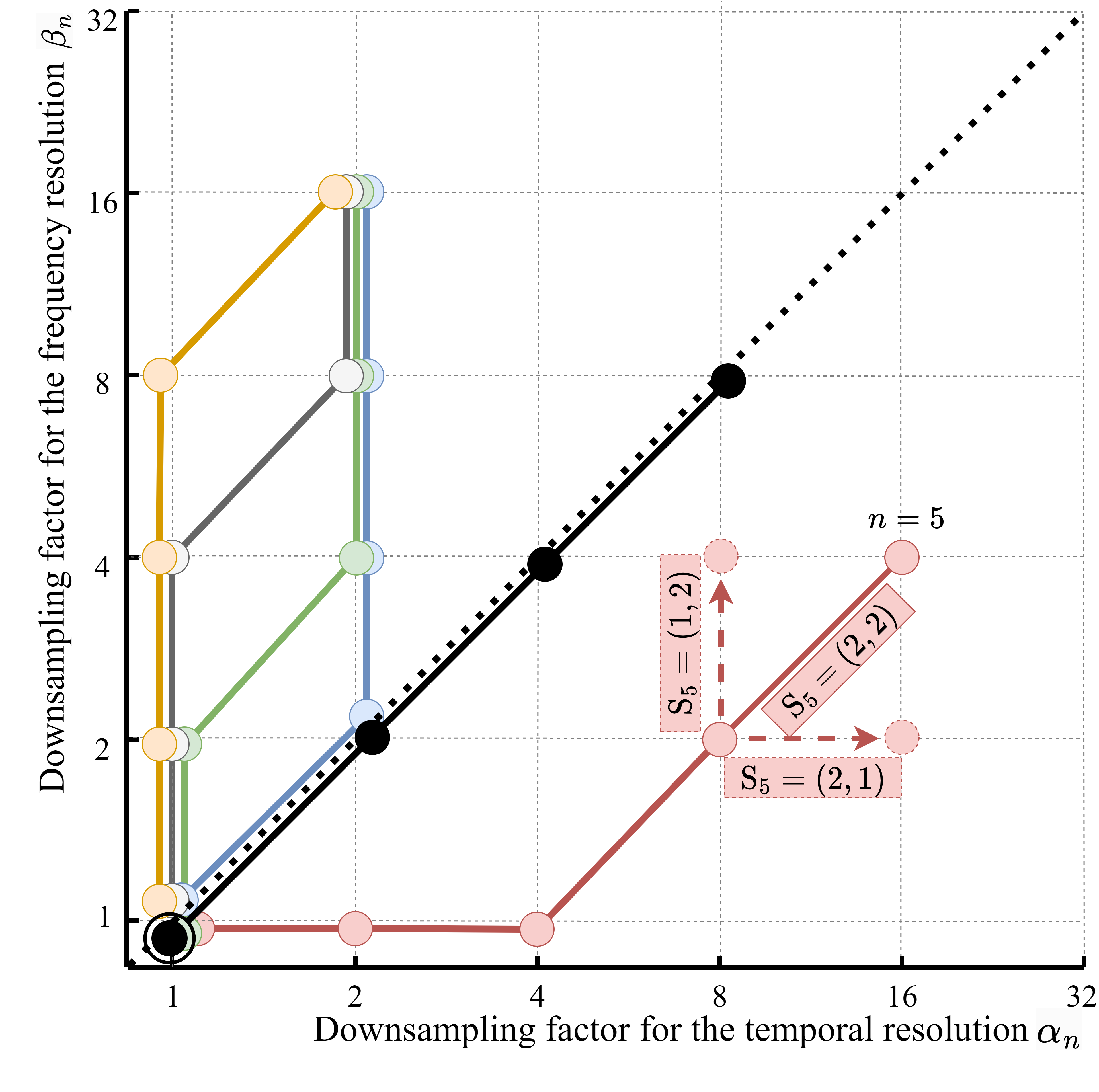}}
\vspace{-4 mm}
\caption{An exemplar trellis diagram. Each node on the trellis diagram represents the time and frequency downsampling factors, $\alpha_{n}$ and $\beta_{n}$, at the output of each stage in a ResNet. Each path represents a stride configuration consisting of five sequential stages, for $n = 1, 2, ..., 5$. The node with a circular outer ring \put(7,2.5){\circle{5}} \put(7,2.5){\circle{8}} \quad~ indicates that it remains at the same position by using a stride of (1,1). Dashed arrows represent two alternative options controlled by different stride operations.}
\label{fig_trellis_example}
\vspace{-2 mm}
\end{figure}

\subsection{Finding the Sweet Spots on the Trellis Diagram}
\label{subsec_Trellis_Diagrams}

To validate the \textbf{\textit{Golden-Gemini Hypothesis}} and to determine the optimal stride configuration for ASV, we carefully design a search strategy in this subsection.
As introduced in Section~\ref{Sec_stride_res_TDNN_resnet}, the temporal and frequency resolutions are primarily controlled by the stride configuration employed on each dimension during the convolution operations. 
In order to visually represent the various stride configurations, we utilize a trellis diagram as a graphical tool to aid our study, as illustrated in Fig.~\ref{fig_trellis_example}. This diagram effectively captures the essence of each stride configuration by illustrating a series of sequential stride operations originating from the start point.

Consider the ResNet structure comprising five stages as detailed in Section~\ref{subsec_resnet} and Table~\ref{table_structure}. 
For {e}ach stride configuration is represented by five sequential steps on the trellis diagram in Fig.~\ref{fig_trellis_example}, with each step denoting a stride operation performed in a ResNet stage.
For the $n$-th stage, the stride operation is denoted as 
$\mathrm{\textbf{S}}_n=(s_{\mathrm{t}, n}, s_{\mathrm{f}, n})$, indicating a reduction in time and frequency resolutions by a pair of stride factor of $s_{\mathrm{t}, n}$ and $s_{\mathrm{f}, n}$, respectively.
When $s_{\mathrm{t}, n}$ or $s_{\mathrm{f}, n}$ equals to 1, the corresponding resolution remains unchanged. 
It's important to highlight that strides of 1 or 2 are the most commonly used and are also the default choices in ResNet~\cite{resnet}. Therefore, in this work, we exclusively focus on these two stride operations.

In Fig.~\ref{fig_trellis_example}, $\alpha_{n}$ and $\beta_{n}$  are the downsampling factors at the output of the $n$-th stage in a ResNet for the temporal and frequency resolutions, respectively.
They are given by the products of the stride factors
, and are formulated as follows.
\begin{equation}
\label{eq_s}
   \alpha_{n} = \prod_{i=1}^{n} s_{\mathrm{t},i},
   \ \mathrm{and} \
   \beta_{n} = \prod_{i=1}^{n} s_{\mathrm{f},i}.
\end{equation}
The output temporal resolution $R_{\mathrm{out, t}, n}$ and frequency resolution $R_{\mathrm{out, f}, n}$ of the $n$-th stage in a ResNet are derived as:

\begin{equation}
\label{eq_time_downsampling}
   R_{\mathrm{out, t}, n} \simeq \frac{R_{\mathrm{init, t}}}{\alpha_{n}}, 
   \ \mathrm{and} \
   R_{\mathrm{out, f}, n} \simeq \frac{R_{\mathrm{init, f}}}{\beta_{n}},
\end{equation}
where $R_{\mathrm{init}}$ is the initial input resolution for the first stage in a ResNet.

For the red path in Fig.~\ref{fig_trellis_example}, a stride of (2,2) reduces both time and frequency resolutions by half at stage $n=5$ simultaneously.
The two dashed arrows indicate the alternative stride configurations for reducing the resolution by half in either the frequency dimension alone with a stride of (1,2), or the time dimension alone with a stride of (2,1).
Additionally, it is allowed to stay at the same node on the trellis diagram, thereby preserving both time and frequency resolutions with a stride of (1,1). This option is denoted as \put(7,2.5){\circle{5}} \put(7,2.5){\circle{8}} \quad~ in Fig.~\ref{fig_trellis_example}, represented by the black node at coordinates (1,1).
The early stages often employ this option to retain sufficient information in low-level features.
This design aligns with that of the modified ResNet~\cite{BUT2019system, wespeaker}, where the first two stages use a stride of = (1,1).

In the trellis diagram depicted in Fig.~\ref{fig_trellis_example}, the stride configurations that prioritize the preservation of either frequency or time resolution are delineated. The endpoints located on the black dotted line indicate stride configurations that treat time and frequency resolutions equally. This black dotted line also divides the diagram into two partitions. Within the upper-left partition, the stride configurations give precedence to the preservation of temporal resolution over frequency resolution. Conversely, the lower-right partition represents configurations that accentuate frequency resolution, while compromising temporal resolution. In our experiments, we search all possible stride configurations within this trellis diagram to identify the optimal stride configuration. The experimental results and analysis of this search are presented in Section~\ref{subsec_eval_Trellis}.

In addition, there are multiple paths on the diagram that lead to a single endpoint, each representing a specific stride configuration. Figure~\ref{fig_trellis_example} shows an exemplar trellis diagram illustrating four paths, each represented by a different color, converging to the endpoint on (2, 16).
The difference among stride configurations that lead to the same endpoint lies in the specific stages within the total of five stages where the downsampling operation with a stride of (2,2) is applied. Performing the downsampling operation in an early stage reduces the resolution of the output feature map, resulting in a smaller feature size that needs to be convolved by 2D convolutions. Consequently, this reduction in resolution contributes to a decrease in FLOPs. Therefore, the stride configurations towards the same endpoint require for different FLOPs while maintaining the same number of parameters. To assess the impact of early or late downsampling, we explored different paths towards the same point with various FLOPs. The experimental results and analysis of the findings are comprehensively presented in Section~\ref{subsec_eval_golden_gemini}.


\section{Experimental Setups}

\subsection{Dataset}

\begin{table}[h]
\begin{center}
\begin{footnotesize}
\caption{Development and Test sets statistics}
\label{table_test_statistics}
\begin{tabular}{crrr}
\hline
\toprule
Test set & \# of speakers & \# of utterances & \# of pairs \\ 
\hline
\midrule
VoxCeleb1-O & 40 & 4,708 & 37,611 \\
VoxCeleb1-H & 1,190 & 137,924 & 550,894 \\
VoxCeleb1-E & 1,251 & 145,160 & 579,818 \\
SITW & - & - & 721,788 \\
CNCeleb & - & - & 3,484,292 \\
\bottomrule
\hline
\end{tabular}
\end{footnotesize}
\end{center}
\end{table}

The experiments are conducted on four large-scale datasets, including the VoxCeleb1~\cite{voxceleb1_dataset}, VoxCeleb2~\cite{voxceleb2_dataset}, \textit{Speaker in the Wild} (SITW)~\cite{SITW_dataset} and CNCeleb~\cite{cnceleb_dataset} datasets.

\textbf{Training set.} During training, only the development partition of the VoxCeleb2 dataset is used, which consists of 5,994 speakers and 1,092,009 utterances. This protocol for training on the VoxCeleb2 dataset is widely adopted~\cite{MFA_TDNN,ECAPA_TDNN, ECAPA_CNN_TDNN, depth_first, depth_first_conf, 10095051, 10023305, pvector2023intspch, wang2023cam++}. Additionally, a randomly selected 2\% portion of this development partition is reserved as the validation set. This small validation set is used to identify the best model for testing on the development and testing sets.

\textbf{Development set.} The VoxCeleb1-Original (Vox1-O) test set is utilized as the development set in this work to conduct a performance comparison of all stride configurations. The outcomes of the tests on this development set are analyzed, leading to the formulation of observations. 

\textbf{Test set.} In order to verify the observations across various scenarios, we comprehensively encompass testing scenarios that include in-domain, out-domain, large-scale, and challenging cases. Specifically, VoxCeleb1-Hard (Vox1-H) and VoxCeleb1-Extended (Vox1-E) are used as in-domain large hard cases and a large test set, respectively. The SITW core-core test set serves the purpose of cross-domain testing, while the CNCeleb test set is employed to assess challenging cases within a cross-domain scenario. The statistics of these four test sets are shown in Table~\ref{table_test_statistics}. It's important to highlight that there is no overlap between any of the test sets and the training set or development set.

\subsection{Training Strategy}
\label{subsec_training_strategy}

The experiments are conducted using the Pytorch framework\footnote{https://pytorch.org/}. We adopted two training strategies as detailed below.

\textbf{Training strategy 1: }
The SpeechBrain Toolkit\footnote{https://speechbrain.github.io/}~\cite{SpeechBrain} is used. For fair comparisons, all systems are trained under the same training strategy following that in~\cite{ECAPA_TDNN, MFA_TDNN}. Specifically, the loss function is the additive angular margin softmax (AAM-softmax)~\cite{arcface} with a margin of 0.2 and a scale of 30. The Adam optimizer~\cite{Adam} with cyclical learning rate~\cite{cyclical} following a triangular policy~\cite{cyclical} is used for training all models. A weight decay of $2 \times 10^{-5}$ is used for all the weights in the model.
The maximum and minimum learning rates of the cyclical scheduler are $2 \times 10^{-3}$ and $2 \times 10^{-8}$, and the batch size is 64 each with 5 types of augmented data. For Res2Net and ResNet101, learning rates and batch size are reduced to half due to the large memory occupation.

All training samples are cut into 3-second segments. 
We employ five augmentation techniques to increase the diversity of the training data. The first two follow the idea of random frame dropout in the time domain~\cite{SpecAugment} and speed perturbation~\cite{speedpur_ASV}. The remaining three are a set of reverberate data, noisy data, and a mixture of both, achieved by combining with the Room Impulse Response (RIR) dataset~\cite{RIR_Dataset}.
The s-norm~\cite{snorm} is applied to normalize the scores.

\textbf{Training strategy 2:}  Wespeaker Toolkit\footnote{https://github.com/wenet-e2e/wespeaker}~\cite{wespeaker} is used. This training strategy follows that in ~\cite{depth_first} for the purpose of re-implementing DF-ResNet~\cite{depth_first} and is only applied to re-implemented DF-ResNet and \textit{Gemini} DF-ResNet reported in Section~\ref{subsec_new_SOTA}. Specifically, the loss function is an AAM-softmax~\cite{arcface} with a margin of 0.2 and a scale of 32. The total number of training epochs is set to 165. The AdamW~\cite{loshchilov2018decoupled} optimizer with 0.05 weight decay is used. 
The base learning rate ($l_{\mathrm{base}}$) decreases from $1.25 \times 10^{-4}$ to $1 \times 10^{-6}$ with the exponential scheduler as the learning rate regulator. The learning rate ($l$) for training is adjusted according to the batch size ($b$) and formulated as $l = l_{\mathrm{base}} \times b / 64$.
 All the samples are cut into 200-frame segments with the augmentations of reverberation, noise, and speed perturbation~\cite{speedpur_ASV} during training. The as-norm~\cite{Matějka2017} is applied to normalize the scores.

\subsection{Evaluation Protocol}
We report the performances in terms of the equal error rate (EER) and the minimum detection cost function (minDCF) with $P_{\mathrm{target}}$ = 0.01 and $C_{\mathrm{FA}}$ = $C_{\mathrm{Miss}}$ = 1.
The scores are produced by calculating the cosine distance between embeddings.

\section{Results and Analysis}
\label{sec_results}
It is worth noting that the FLOPs calculation is correlated with the duration of the sample. We select the most commonly used options of 2 seconds~\cite{depth_first, depth_first_conf, wespeaker, ECAPA_CNN_TDNN,ECAPA_TDNN, 10023305} and 3 seconds~\cite{MFA_TDNN,SpeechBrain, zhang22h_interspeech, wang2023cam++, 9746050, 9747384, QQ2023incorporating} to calculate FLOPs. The results are labeled as `2s/3s'.

\subsection{Original ResNet v.s. Modified ResNet (Baseline)}
\label{subsec_eval_ori_vs_mod}

We first compare the modified ResNet~\cite{BUT2019system} and original ResNet~\cite{resnet}. The results are presented in Table~\ref{Table_ori_mod}. It is obvious that the modified ResNet outperforms the original ResNet.
The improved performance of the modified ResNet is attributed to the adequate preservation of frequency-time resolution by changing the stride configurations from (2,2) to (1,1) in the first two layers. However, these changes also lead to an increase in FLOPs.

In addition, as this modified ResNet~\cite{BUT2019system} achieves SOTA performance using the equal-stride configuration, we adopt it as the baseline model in this work.

\begin{table}[t]
\caption{Performance in EER(\%) and minDCF of original ResNet~\cite{resnet} and modified ResNet~\cite{BUT2019system, wespeaker}. The FLOPs are calculated based on a 3-second sample.}
\vspace{-2 mm}
\label{Table_ori_mod}
\begin{center}
\begin{scriptsize}
\setlength{\tabcolsep}{0.5mm}{
\renewcommand{\arraystretch}{1.05}{
\begin{tabular}{cccccccc}
\hline
\toprule
 &  &  & Vox1-O & Vox1-H & Vox1-E & SITW & CNCeleb \\ \cline{4-8} 
 &  &  & EER & EER & EER & EER & EER \\
\multirow{-3}{*}{Model} & \multirow{-3}{*}{\begin{tabular}[c]{@{}c@{}}Params\\ (Million)\end{tabular}} & \multirow{-3}{*}{\begin{tabular}[c]{@{}c@{}}FLOPs\\ (Giga)\end{tabular}} & minDCF & minDCF & minDCF & minDCF & minDCF \\

\hline
\midrule

\rowcolor[HTML]{EFEFEF} 
\cellcolor[HTML]{EFEFEF} & \cellcolor[HTML]{EFEFEF} & \cellcolor[HTML]{EFEFEF} & 2.903 & 5.106 & 2.934 & 3.609 & 16.373 \\
\rowcolor[HTML]{EFEFEF} 
\multirow{-2}{*}{\cellcolor[HTML]{EFEFEF}original ResNet18} & \multirow{-2}{*}{\cellcolor[HTML]{EFEFEF}11.3} & \multirow{-2}{*}{\cellcolor[HTML]{EFEFEF}\textbf{0.90}} & 0.315 & 0.433 & 0.322 & 0.396 & 1.000 \\
 &  &  & \textbf{1.760} & \textbf{2.785} & \textbf{1.600} & \textbf{2.132} & \textbf{12.301} \\
\multirow{-2}{*}{modified ResNet18} & \multirow{-2}{*}{\textbf{3.45}} & \multirow{-2}{*}{3.25} & \textbf{0.177} & \textbf{0.244} & \textbf{0.170} & \textbf{0.210} & \textbf{0.657} \\ \hline
\rowcolor[HTML]{EFEFEF} 
\cellcolor[HTML]{EFEFEF} & \cellcolor[HTML]{EFEFEF} & \cellcolor[HTML]{EFEFEF} & 2.744 & 4.598 & 2.614 & 3.308 & 15.072 \\
\rowcolor[HTML]{EFEFEF} 
\multirow{-2}{*}{\cellcolor[HTML]{EFEFEF}original ResNet34} & \multirow{-2}{*}{\cellcolor[HTML]{EFEFEF}21.41} & \multirow{-2}{*}{\cellcolor[HTML]{EFEFEF}\textbf{1.82}} & 0.291 & 0.400 & 0.284 & 0.379 & 1.000 \\
 &  &  & \textbf{1.101} & \textbf{2.221} & \textbf{1.252} & \textbf{1.584} & \textbf{12.113} \\
\multirow{-2}{*}{modified ResNet34} & \multirow{-2}{*}{\textbf{6.63}} & \multirow{-2}{*}{6.88} & \textbf{0.128} & \textbf{0.208} & \textbf{0.139} & \textbf{0.161} & \textbf{0.623} \\

\bottomrule
\hline
\end{tabular}}}
\end{scriptsize}
\end{center}
\vspace{-3 mm}
\end{table}

\subsection{Finding the Sweet Spots on the Trellis Diagram}
\label{subsec_eval_Trellis}

\begin{table*}[t]
\caption{Performance in EER(\%) and minDCF of the original ResNet34 (ORI)~\cite{resnet} and the modified ResNet34 (MOD)~\cite{BUT2019system, wespeaker} with different stride configurations demonstrated in Fig.~\ref{fig_Trellis_diagram} (a). Experiments are conducted on the development set (Vox1-O) in the left sub-table and on the test sets (Vox1-H, Vox1-E, SITW, CNCeleb) in the right sub-table.
The stride configuration shows the stride factors for time and frequency dimensions in the five stages of ResNet architecture.
The symbol ↑↓ indicates the average relative changes across all four test sets over the benchmark model.}
\label{Table_all}
\begin{minipage}{0.53\linewidth}
\begin{center}
\begin{footnotesize}
\renewcommand\arraystretch{1.07}
\setlength{\tabcolsep}{1mm}{
\begin{tabular}{cccccc}
\hline
\toprule
 & Downsampling & Stride Config. &  &  & Vox1-O \\ \cline{3-3} \cline{6-6} 
 & Factors & [Time] &  & \multirow{-2}{*}{\begin{tabular}[c]{@{}c@{}}FLOPs\\ (Giga)\end{tabular}} & EER \\ \cline{2-2} \cline{5-5}
\multirow{-3}{*}{\begin{tabular}[c]{@{}c@{}}Index of \\ Stride \\ Config.\end{tabular}} & ($\alpha_{\mathrm{5}}$, $\beta_{\mathrm{5}}$) & [Frequency] & \multirow{-3}{*}{\begin{tabular}[c]{@{}c@{}}Params\\ (Million)\end{tabular}} & 2s/3s & minDCF \\ 

\hline
\midrule

\rowcolor[HTML]{EFEFEF} 
\cellcolor[HTML]{EFEFEF} & \cellcolor[HTML]{EFEFEF} & {[}2,2,2,2,2{]} & \cellcolor[HTML]{EFEFEF} & \cellcolor[HTML]{EFEFEF} & 2.744\\
\rowcolor[HTML]{EFEFEF} 
\multirow{-2}{*}{\cellcolor[HTML]{EFEFEF}ORI} & \multirow{-2}{*}{\cellcolor[HTML]{EFEFEF}(32,32)} & {[}2,2,2,2,2{]} & \multirow{-2}{*}{\cellcolor[HTML]{EFEFEF}21.41} & \multirow{-2}{*}{\cellcolor[HTML]{EFEFEF}1.25/1.82} & 0.291 \\
 &  & {[}1,1,2,2,2{]} &  &  & 1.101 \\
\multirow{-2}{*}{MOD} & \multirow{-2}{*}{(8,8)} & {[}1,1,2,2,2{]} & \multirow{-2}{*}{6.63} & \multirow{-2}{*}{4.63/6.88} & 0.128 \\ \hline
\rowcolor[HTML]{EFEFEF} 
\cellcolor[HTML]{EFEFEF} & \cellcolor[HTML]{EFEFEF} & {[}1,1,1,1,1{]} & \cellcolor[HTML]{EFEFEF} & \cellcolor[HTML]{EFEFEF} & 1.250\\
\rowcolor[HTML]{EFEFEF} 
\multirow{-2}{*}{\cellcolor[HTML]{EFEFEF}T05} & \multirow{-2}{*}{\cellcolor[HTML]{EFEFEF}(1,32)} & {[}2,2,2,2,2{]} & \multirow{-2}{*}{\cellcolor[HTML]{EFEFEF}5.72} & \multirow{-2}{*}{\cellcolor[HTML]{EFEFEF}4.49/6.72} & 0.131 \\
 &  & {[}2,2,2,2,2{]} &  &  & 2.526\\
\multirow{-2}{*}{F50} & \multirow{-2}{*}{(32,1)} & {[}1,1,1,1,1{]} & \multirow{-2}{*}{15.81} & \multirow{-2}{*}{4.44/6.43} & 0.241  \\
\rowcolor[HTML]{EFEFEF} 
\cellcolor[HTML]{EFEFEF} & \cellcolor[HTML]{EFEFEF} & \cellcolor[HTML]{EFEFEF}{[}1,1,1,1,2{]} & \cellcolor[HTML]{EFEFEF} & \cellcolor[HTML]{EFEFEF} & 1.303 \\
\rowcolor[HTML]{EFEFEF} 
\multirow{-2}{*}{\cellcolor[HTML]{EFEFEF}T15} & \multirow{-2}{*}{\cellcolor[HTML]{EFEFEF}(2,32)} & \cellcolor[HTML]{EFEFEF}{[}2,2,2,2,2{]} & \multirow{-2}{*}{\cellcolor[HTML]{EFEFEF}5.72} & \multirow{-2}{*}{\cellcolor[HTML]{EFEFEF}3.50/5.24} & 0.161  \\
 &  & {[}2,2,2,2,2{]} &  &  & 2.505 \\
\multirow{-2}{*}{F51} & \multirow{-2}{*}{(32,2)} & {[}1,1,1,1,2{]} & \multirow{-2}{*}{10.57} & \multirow{-2}{*}{3.52/5.11} & 0.243  \\
\rowcolor[HTML]{EFEFEF} 
\cellcolor[HTML]{EFEFEF} & \cellcolor[HTML]{EFEFEF} & {[}1,1,1,2,2{]} & \cellcolor[HTML]{EFEFEF} & \cellcolor[HTML]{EFEFEF} & 1.218  \\
\rowcolor[HTML]{EFEFEF} 
\multirow{-2}{*}{\cellcolor[HTML]{EFEFEF}T25} & \multirow{-2}{*}{\cellcolor[HTML]{EFEFEF}(4,32)} & {[}2,2,2,2,2{]} & \multirow{-2}{*}{\cellcolor[HTML]{EFEFEF}5.72} & \multirow{-2}{*}{\cellcolor[HTML]{EFEFEF}2.16/3.23} & 0.133  \\
 &  & {[}2,2,2,2,2{]} &  &  & 2.228 \\
\multirow{-2}{*}{F52} & \multirow{-2}{*}{(32,4)} & {[}1,1,1,2,2{]} & \multirow{-2}{*}{7.95} & \multirow{-2}{*}{2.17/3.16} & 0.219  \\
\rowcolor[HTML]{EFEFEF} 
\cellcolor[HTML]{EFEFEF} & \cellcolor[HTML]{EFEFEF} & {[}1,1,1,1,2{]} & \cellcolor[HTML]{EFEFEF} & \cellcolor[HTML]{EFEFEF} & \textbf{1.058}  \\
\rowcolor[HTML]{EFEFEF} 
\multirow{-2}{*}{\cellcolor[HTML]{EFEFEF}T14} & \multirow{-2}{*}{\cellcolor[HTML]{EFEFEF}(2,16)} & {[}1,2,2,2,2{]} & \multirow{-2}{*}{\cellcolor[HTML]{EFEFEF}5.98} & \multirow{-2}{*}{\cellcolor[HTML]{EFEFEF}6.68/9.99} & \textbf{0.092} \\
 &  & {[}1,2,2,2,2{]} &  &  & 1.882  \\
\multirow{-2}{*}{F41} & \multirow{-2}{*}{(16,2)} & {[}1,1,1,1,2{]} & \multirow{-2}{*}{10.57} & \multirow{-2}{*}{6.87/10.08} & 0.150  \\
\rowcolor[HTML]{EFEFEF} 
\cellcolor[HTML]{EFEFEF} & \cellcolor[HTML]{EFEFEF} & {[}1,1,1,2,2{]} & \cellcolor[HTML]{EFEFEF} & \cellcolor[HTML]{EFEFEF} & 1.111 \\
\rowcolor[HTML]{EFEFEF} 
\multirow{-2}{*}{\cellcolor[HTML]{EFEFEF}T24} & \multirow{-2}{*}{\cellcolor[HTML]{EFEFEF}(4,16)} & {[}1,2,2,2,2{]} & \multirow{-2}{*}{\cellcolor[HTML]{EFEFEF}5.98} & \multirow{-2}{*}{\cellcolor[HTML]{EFEFEF}4.15/6.20} & 0.104  \\
 &  & {[}1,2,2,2,2{]} &  &  & 1.563  \\
\multirow{-2}{*}{F42} & \multirow{-2}{*}{(16,4)} & {[}1,1,1,2,2{]} & \multirow{-2}{*}{7.95} & \multirow{-2}{*}{4.24/6.24} & 0.135  \\
\rowcolor[HTML]{EFEFEF} 
\cellcolor[HTML]{EFEFEF} & \cellcolor[HTML]{EFEFEF} & {[}1,1,2,2,2{]} & \cellcolor[HTML]{EFEFEF} & \cellcolor[HTML]{EFEFEF} & 1.260  \\
\rowcolor[HTML]{EFEFEF} 
\multirow{-2}{*}{\cellcolor[HTML]{EFEFEF}T34} & \multirow{-2}{*}{\cellcolor[HTML]{EFEFEF}(8,16)} & {[}1,2,2,2,2{]} & \multirow{-2}{*}{\cellcolor[HTML]{EFEFEF}5.98} & \multirow{-2}{*}{\cellcolor[HTML]{EFEFEF}2.32/3.46} & 0.128  \\
 &  & {[}1,2,2,2,2{]} &  &  & 1.691 \\
\multirow{-2}{*}{F43} & \multirow{-2}{*}{(16,8)} & {[}1,1,2,2,2{]} & \multirow{-2}{*}{6.64} & \multirow{-2}{*}{2.35/3.47} & 0.180\\
\rowcolor[HTML]{EFEFEF} 
\cellcolor[HTML]{EFEFEF} & \cellcolor[HTML]{EFEFEF} & {[}1,1,1,2,2{]} & \cellcolor[HTML]{EFEFEF} & \cellcolor[HTML]{EFEFEF} & 1.101  \\
\rowcolor[HTML]{EFEFEF} 
\multirow{-2}{*}{\cellcolor[HTML]{EFEFEF}T23} & \multirow{-2}{*}{\cellcolor[HTML]{EFEFEF}(4,8)} & {[}1,1,2,2,2{]} & \multirow{-2}{*}{\cellcolor[HTML]{EFEFEF}6.63} & \multirow{-2}{*}{\cellcolor[HTML]{EFEFEF}8.27/12.37} & 0.095  \\
 &  & {[}1,1,2,2,2{]} &  &  & 1.223 \\
\multirow{-2}{*}{F32} & \multirow{-2}{*}{(8,4)} & {[}1,1,1,2,2{]} & \multirow{-2}{*}{7.95} & \multirow{-2}{*}{8.35/12.41} & 0.105  \\ 
\rowcolor[HTML]{EFEFEF} 
\cellcolor[HTML]{EFEFEF} & \cellcolor[HTML]{EFEFEF} & {[}1,1,1,1,1{]} & \cellcolor[HTML]{EFEFEF} & \cellcolor[HTML]{EFEFEF} & 1.276  \\
\rowcolor[HTML]{EFEFEF} 
\multirow{-2}{*}{\cellcolor[HTML]{EFEFEF}T04} & \multirow{-2}{*}{\cellcolor[HTML]{EFEFEF}(1,16)} & {[}1,2,2,2,2{]} & \multirow{-2}{*}{\cellcolor[HTML]{EFEFEF}5.98} & \multirow{-2}{*}{\cellcolor[HTML]{EFEFEF}8.32/12.45} & 0.117  \\
 &  & {[}1,1,1,1,2{]} &  &  & 1.127  \\
\multirow{-2}{*}{T13} & \multirow{-2}{*}{(2,8)} & {[}1,1,2,2,2{]} & \multirow{-2}{*}{6.63} & \multirow{-2}{*}{13.33/19.95} & 0.107 \\ 

\bottomrule
\hline
\end{tabular}
}
\end{footnotesize}
\end{center}
\end{minipage}\quad \quad
\begin{minipage}{0.43\linewidth}
\begin{center}
\begin{footnotesize}
\renewcommand\arraystretch{1.07}
\setlength{\tabcolsep}{1mm}{
\begin{tabular}{cccccc}
\hline
\toprule

 &  Vox1-H & Vox1-E & SITW & CNCeleb & ↑↓ \\  \cline{2-6} 
 & EER & EER & EER & EER & EER \\  
\multirow{-3}{*}{\begin{tabular}[c]{@{}c@{}}Index of \\ Stride \\ Config.\end{tabular}} & minDCF & minDCF & minDCF & minDCF & minDCF \\ 

\hline
\midrule

\rowcolor[HTML]{EFEFEF} 
\cellcolor[HTML]{EFEFEF} & 4.598 & 2.614 & 3.308 & 15.072 & +87.27\% \\
\rowcolor[HTML]{EFEFEF} 
\multirow{-2}{*}{\cellcolor[HTML]{EFEFEF}ORI} & 0.400 & 0.284 & 0.379 & 1.000 & +98.30\% \\
  & 2.221 & 1.252 & 1.584 & 12.113 & Benchmark \\
\multirow{-2}{*}{MOD} & 0.208 & 0.139 & 0.161 & 0.623 & Benchmark \\ \hline
\rowcolor[HTML]{EFEFEF} 
\cellcolor[HTML]{EFEFEF}  & 2.297 & 1.359 & 1.914 & 12.149 & \cellcolor[HTML]{EFEFEF}+8.27\% \\
\rowcolor[HTML]{EFEFEF} 
\multirow{-2}{*}{\cellcolor[HTML]{EFEFEF}T05} & 0.211 & 0.141 & 0.177 & 0.587 & \cellcolor[HTML]{EFEFEF}+1.79\% \\
 & 3.932 & 2.361 & 3.262 & 12.898 & +69.51\% \\
\multirow{-2}{*}{F50} & 0.329 & 0.249 & 0.316 & 0.755 & +63.83\% \\
\rowcolor[HTML]{EFEFEF} 
\cellcolor[HTML]{EFEFEF} & 2.251 & 1.319 & 1.832 & 12.115 & +5.60\% \\
\rowcolor[HTML]{EFEFEF} 
\multirow{-2}{*}{\cellcolor[HTML]{EFEFEF}T15} & 0.210 & 0.137 & 0.172 & 0.626 & +1.54\% \\
  & 3.914 & 2.350 & 3.216 & 12.678 & +67.91\% \\
\multirow{-2}{*}{F51}  & 0.332 & 0.245 & 0.314 & 0.797 & +64.58\% \\
\rowcolor[HTML]{EFEFEF} 
\cellcolor[HTML]{EFEFEF} & 2.266 & 1.282 & 1.640 & 13.027 & \cellcolor[HTML]{EFEFEF}+3.88\% \\
\rowcolor[HTML]{EFEFEF} 
\multirow{-2}{*}{\cellcolor[HTML]{EFEFEF}T25}  & 0.212 & 0.138 & 0.174 & 0.680 & \cellcolor[HTML]{EFEFEF}+4.61\% \\
  & 3.720 & 2.195 & 2.925 & 12.937 & +58.58\% \\
\multirow{-2}{*}{F52}  & 0.330 & 0.235 & 0.280 & 0.804 & +57.83\% \\
\rowcolor[HTML]{EFEFEF} 
\cellcolor[HTML]{EFEFEF}  & 1.998 & 1.148 & 1.505 & 11.670 & \cellcolor[HTML]{EFEFEF}-6.75\% \\
\rowcolor[HTML]{EFEFEF} 
\multirow{-2}{*}{\cellcolor[HTML]{EFEFEF}T14} & 0.185 & \textbf{0.120} & 0.148 & 0.549 & \cellcolor[HTML]{EFEFEF}-11.14\% \\
  & 2.982 & 1.766 & 2.433 & 12.222 & +32.46\% \\
\multirow{-2}{*}{F41}  & 0.256 & 0.185 & 0.240 & 0.667 & +28.11\% \\
\rowcolor[HTML]{EFEFEF} 
\cellcolor[HTML]{EFEFEF}  & 2.084 & 1.177 & 1.558 & 12.008 & \cellcolor[HTML]{EFEFEF}-3.66\% \\
\rowcolor[HTML]{EFEFEF} 
\multirow{-2}{*}{\cellcolor[HTML]{EFEFEF}T24}  & 0.193 & 0.125 & 0.149 & 0.607 & \cellcolor[HTML]{EFEFEF}-6.80\% \\
 & 2.765 & 1.577 & 2.105 & 11.878 & +20.35\% \\
\multirow{-2}{*}{F42}  & 0.245 & 0.178 & 0.227 & 0.768 & +27.52\% \\
\rowcolor[HTML]{EFEFEF} 
\cellcolor[HTML]{EFEFEF} & 2.392 & 1.328 & 1.750 & 12.487 & \cellcolor[HTML]{EFEFEF}+6.83\% \\
\rowcolor[HTML]{EFEFEF} 
\multirow{-2}{*}{\cellcolor[HTML]{EFEFEF}T34}  & 0.222 & 0.147 & 0.169 & 0.705 & \cellcolor[HTML]{EFEFEF}+7.75\% \\
  & 2.916 & 1.636 & 2.132 & 12.504 & +24.95\% \\
\multirow{-2}{*}{F43} & 0.266 & 0.178 & 0.227 & 0.864 & +33.93\% \\
\rowcolor[HTML]{EFEFEF} 
\cellcolor[HTML]{EFEFEF} & \textbf{1.965} & \textbf{1.110} & \textbf{1.394} & \textbf{11.141} & \cellcolor[HTML]{EFEFEF}-10.72\% \\
\rowcolor[HTML]{EFEFEF} 
\multirow{-2}{*}{\cellcolor[HTML]{EFEFEF}T23} & \textbf{0.182} & 0.121 & \textbf{0.140} & 0.572 & \cellcolor[HTML]{EFEFEF}-11.62\% \\
  & 2.124 & 1.217 & 1.476 & 12.034 & -3.65\% \\
\multirow{-2}{*}{F32} & 0.199 & 0.131 & 0.161 & 0.621 & -2.61\% \\ 
\rowcolor[HTML]{EFEFEF} 
\cellcolor[HTML]{EFEFEF} & 2.182 & 1.282 & 1.777 & 11.484 & \cellcolor[HTML]{EFEFEF}+1.92\% \\
\rowcolor[HTML]{EFEFEF} 
\multirow{-2}{*}{\cellcolor[HTML]{EFEFEF}T04}  & 0.197 & 0.133 & 0.170 & 0.600 & \cellcolor[HTML]{EFEFEF}-1.94\% \\
  & 2.009 & 1.160 & 1.563 & 11.400 & -6.02\% \\
\multirow{-2}{*}{T13} & 0.183 & 0.121 & 0.144 & \textbf{0.548} & -11.87\% \\

\bottomrule
\hline
\end{tabular}
}
\end{footnotesize}
\end{center}
\end{minipage}

\vskip -0.10in
\end{table*}


\begin{figure*}[t]
\centerline{\includegraphics[scale=0.066]{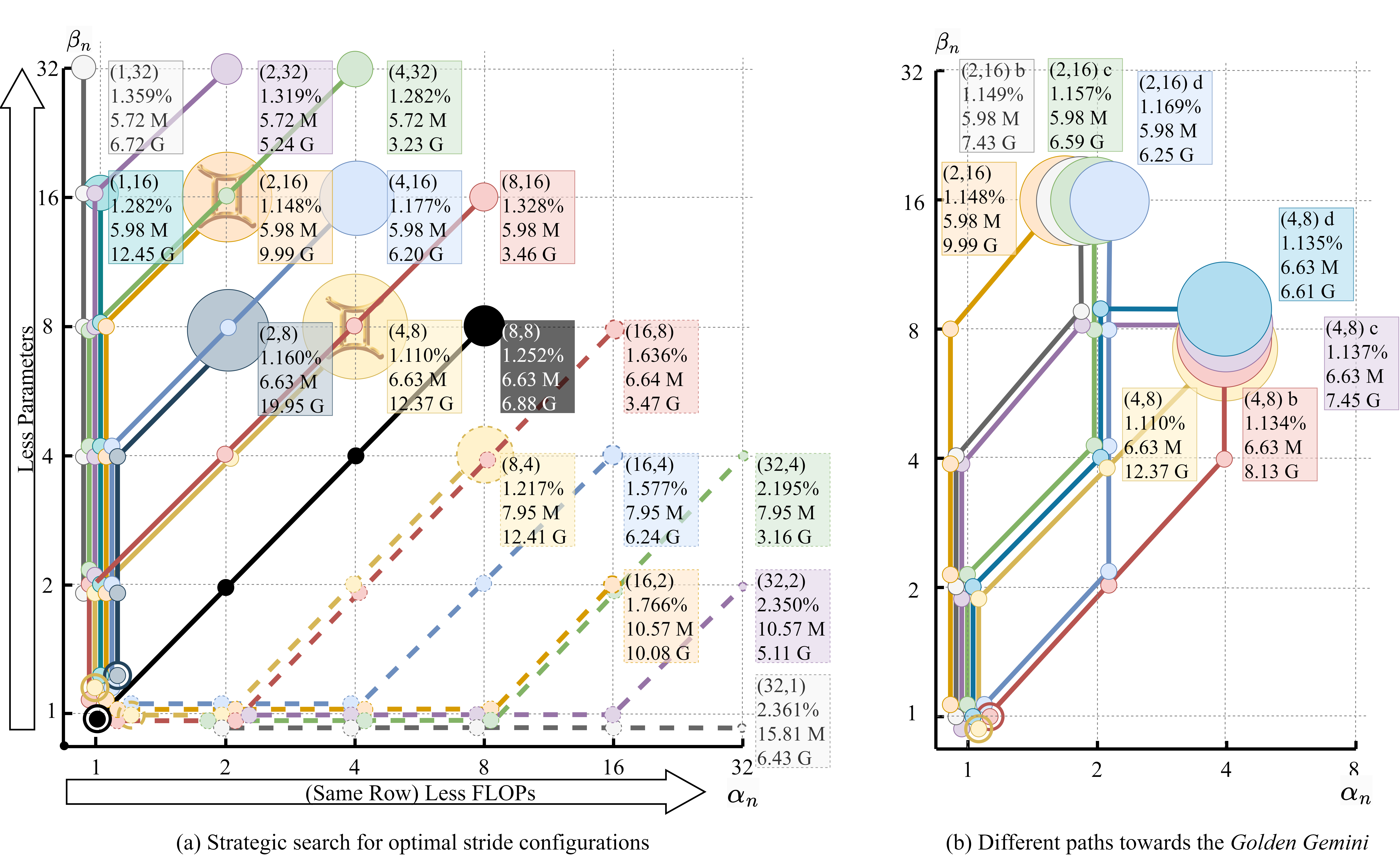}}
\vspace{-1 mm}
\caption{Trellis diagrams of (a) the strategic search for optimal stride configurations and (b) different paths towards  \textit{Golden Gemini}.  \textbf{\textcolor{Bittersweet}{$\gemini$}}  in (a) indicates proposed \textit{Golden-Gemini} stride configurations. In the rectangle box, from top to bottom are: the downsampling factors ($\alpha_5, \beta_5$), performance in EER (\%) on VoxCeleb-E test set, number of parameters, and FLOPs. The size of the endpoint bubble indicates the performance, and the larger the bubble, the better the performance. The node with a circular outer ring forming as \put(7,2.5){\circle{5}} \put(7,2.5){\circle{8}} \quad~ indicates that it remains at the same position by using a stride of (1,1). The solid line represents a stride configuration that prioritizes temporal resolution over frequency resolution, while the dashed line configuration reflects the opposite.}
\label{fig_Trellis_diagram}
\vspace{-2 mm}
\end{figure*}

\begin{figure}[t]
\centerline{\includegraphics[scale=0.062]{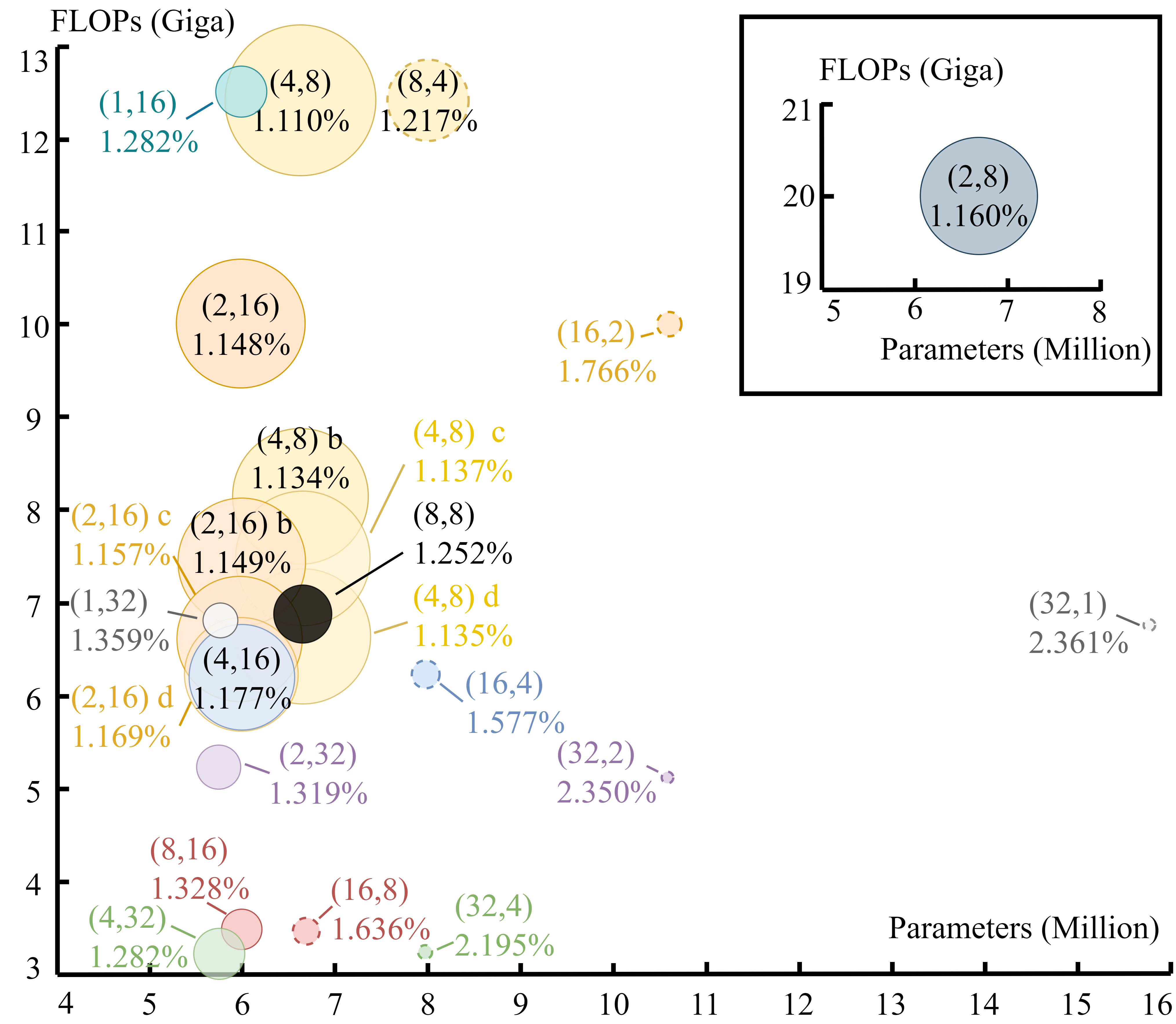}}
\caption{Performance versus FLOPs and the number of parameters for different stride configurations in Fig.~\ref{fig_Trellis_diagram}. The color is consistent with Fig.~\ref{fig_Trellis_diagram} (a). The size of the bubble indicates the performance in EER (\%) on the VoxCeleb-E test set, and the larger the bubble, the better the performance.}
\label{fig_compare_all_stride}
\vspace{-6 mm}
\end{figure}

\begin{table*}[]
\caption{Performance in EER(\%) and minDCF of the modified ResNet34~\cite{BUT2019system,resnet} and \textit{Golden Gemini} models with different paths demonstrated in Fig.~\ref{fig_Trellis_diagram} (b). Experiments are conducted on the development set (Vox1-O) in the left sub-table and on the test sets (Vox1-H, Vox1-E, SITW, CNCeleb) in the right sub-table.
The stride configuration shows the stride factors for time and frequency dimensions in the five stages of ResNet architecture.
The symbol ↑↓ indicates the average relative changes across all four test sets compared to the benchmark model.}
\label{table_Golden_Gemini}

\begin{minipage}{0.53\linewidth}
\begin{center}
\begin{footnotesize}
\renewcommand\arraystretch{1.03}
\setlength{\tabcolsep}{1mm}{
\begin{tabular}{cccccc}

\hline
\toprule
 & Downsampling & Stride Config. &  &  & Vox1-O \\ \cline{3-3} \cline{6-6} 
 & Factors & [Time] &  & \multirow{-2}{*}{\begin{tabular}[c]{@{}c@{}}FLOPs\\ (Giga)\end{tabular}} & EER \\ \cline{2-2} \cline{5-5}
\multirow{-3}{*}{\begin{tabular}[c]{@{}c@{}}Index of \\ Stride \\ Config.\end{tabular}} & ($\alpha_{\mathrm{5}}$, $\beta_{\mathrm{5}}$) & [Frequency] & \multirow{-3}{*}{\begin{tabular}[c]{@{}c@{}}Params\\ (Million)\end{tabular}} & 2s/3s & minDCF \\

\midrule

\rowcolor[HTML]{EFEFEF} 
\cellcolor[HTML]{EFEFEF} & \cellcolor[HTML]{EFEFEF} & {[}1,1,2,2,2{]} & \cellcolor[HTML]{EFEFEF} & \cellcolor[HTML]{EFEFEF} & 1.101 \\
\rowcolor[HTML]{EFEFEF} 
\multirow{-2}{*}{\cellcolor[HTML]{EFEFEF}MOD} & \multirow{-2}{*}{\cellcolor[HTML]{EFEFEF}(8,8)} & {[}1,1,2,2,2{]} & \multirow{-2}{*}{\cellcolor[HTML]{EFEFEF}6.63} & \multirow{-2}{*}{\cellcolor[HTML]{EFEFEF}4.63/6.88} & 0.128 \\ \hline
 &  & {[}1,1,1,1,2{]} &  &  & 1.058  \\
\multirow{-2}{*}{T14} &  & {[}1,2,2,2,2{]} & \multirow{-2}{*}{5.98} & \multirow{-2}{*}{6.68/9.99} & \textbf{0.092} \\
\cellcolor[HTML]{EFEFEF} &  & \cellcolor[HTML]{EFEFEF}{[}1,1,1,2,1{]} & \cellcolor[HTML]{EFEFEF} & \cellcolor[HTML]{EFEFEF} & \cellcolor[HTML]{EFEFEF}1.056  \\
\multirow{-2}{*}{\cellcolor[HTML]{EFEFEF}T14b} &  & \cellcolor[HTML]{EFEFEF}{[}{1},2,2,2,2{]} & \multirow{-2}{*}{\cellcolor[HTML]{EFEFEF}5.98} & \multirow{-2}{*}{\cellcolor[HTML]{EFEFEF}4.97/7.43} & \cellcolor[HTML]{EFEFEF}0.093 \\
 &  & {[}1,1,2,1,1{]} &  &  & \textbf{1.053} \\
\multirow{-2}{*}{T14c} &  & {[}1,2,2,2,2{]} & \multirow{-2}{*}{5.98} & \multirow{-2}{*}{4.41/6.59} & \textbf{0.092} \\
\cellcolor[HTML]{EFEFEF} &  & \cellcolor[HTML]{EFEFEF}{[}1,2,1,1,1{]} & \cellcolor[HTML]{EFEFEF} & \cellcolor[HTML]{EFEFEF} & \cellcolor[HTML]{EFEFEF}1.154 \\
\multirow{-2}{*}{\cellcolor[HTML]{EFEFEF}T14d} & \multirow{-8}{*}{(2,16)} & \cellcolor[HTML]{EFEFEF}{[}{1},2,2,2,2{]} & \multirow{-2}{*}{\cellcolor[HTML]{EFEFEF}5.98} & \multirow{-2}{*}{\cellcolor[HTML]{EFEFEF}4.18/6.25} & \cellcolor[HTML]{EFEFEF}0.115  \\ \hline
 &  & {[}1,1,1,2,2{]} &  &  & 1.101  \\
\multirow{-2}{*}{T23} &  & {[}1,1,2,2,2{]} & \multirow{-2}{*}{6.63} & \multirow{-2}{*}{8.27/12.37} & 0.095  \\
\cellcolor[HTML]{EFEFEF} &  & \cellcolor[HTML]{EFEFEF}{[}1,1,2,2,1{]} & \cellcolor[HTML]{EFEFEF} & \cellcolor[HTML]{EFEFEF} & \cellcolor[HTML]{EFEFEF}1.095  \\
\multirow{-2}{*}{\cellcolor[HTML]{EFEFEF}T23b} &  & \cellcolor[HTML]{EFEFEF}{[}1,1,2,2,2{]} & \multirow{-2}{*}{\cellcolor[HTML]{EFEFEF}6.63} & \multirow{-2}{*}{\cellcolor[HTML]{EFEFEF}5.45/8.13} & \cellcolor[HTML]{EFEFEF}0.099 \\
 &  & {[}1,1,1,2,2{]} &  &  & 1.122  \\
\multirow{-2}{*}{T23c} &  & {[}1,2,2,2,1{]} & \multirow{-2}{*}{6.63} & \multirow{-2}{*}{4.99/7.45} & 0.112 \\
\cellcolor[HTML]{EFEFEF} &  & \cellcolor[HTML]{EFEFEF}{[}1,1,2,1,2{]} & \cellcolor[HTML]{EFEFEF} & \cellcolor[HTML]{EFEFEF} & \cellcolor[HTML]{EFEFEF}1.095  \\
\multirow{-2}{*}{\cellcolor[HTML]{EFEFEF}T23d} & \multirow{-8}{*}{(4,8)} & \cellcolor[HTML]{EFEFEF}{[}1,2,2,2,1{]} & \multirow{-2}{*}{\cellcolor[HTML]{EFEFEF}6.63} & \multirow{-2}{*}{\cellcolor[HTML]{EFEFEF}4.43/6.61} & \cellcolor[HTML]{EFEFEF}0.104  \\

\bottomrule
\hline
\end{tabular}
}
\end{footnotesize}
\end{center}
\end{minipage}\quad \quad
\begin{minipage}{0.43\linewidth}
\begin{center}
\begin{footnotesize}
\renewcommand\arraystretch{1.03}
\setlength{\tabcolsep}{1.1mm}{
\begin{tabular}{cccccc}

\hline
\toprule

 &  Vox1-H & Vox1-E & SITW & CNCeleb & ↑↓ \\  \cline{2-6} 
 & EER & EER & EER & EER & EER \\  
\multirow{-3}{*}{\begin{tabular}[c]{@{}c@{}}Index of \\ Stride \\ Config.\end{tabular}} & minDCF & minDCF & minDCF & minDCF & minDCF \\

\midrule

\rowcolor[HTML]{EFEFEF} 
\cellcolor[HTML]{EFEFEF} & 2.221 & 1.252 & 1.584 & 12.113 & Benchmark \\
\rowcolor[HTML]{EFEFEF} 
\multirow{-2}{*}{\cellcolor[HTML]{EFEFEF}MOD}  & 0.208 & 0.139 & 0.161 & 0.623 & Benchmark \\ \hline
  & 1.998 & 1.148 & 1.505 & 11.670 & -6.75\% \\
\multirow{-2}{*}{T14}  & 0.185 & 0.120 & 0.148 & 0.549 & -11.14\% \\
\cellcolor[HTML]{EFEFEF}  & \cellcolor[HTML]{EFEFEF}2.023 & \cellcolor[HTML]{EFEFEF}1.149 & \cellcolor[HTML]{EFEFEF}1.531 & \cellcolor[HTML]{EFEFEF}11.715 & \cellcolor[HTML]{EFEFEF}-5.94\% \\
\multirow{-2}{*}{\cellcolor[HTML]{EFEFEF}T14b}  & \cellcolor[HTML]{EFEFEF}0.190 & \cellcolor[HTML]{EFEFEF}0.124 & \cellcolor[HTML]{EFEFEF}0.151 & \cellcolor[HTML]{EFEFEF}0.559 & \cellcolor[HTML]{EFEFEF}-9.07\% \\
  & 2.010 & 1.157 & 1.504 & 11.828 & -6.13\% \\
\multirow{-2}{*}{T14c}  & 0.186 & 0.124 & 0.146 & \textbf{0.540} & -10.99\% \\
\cellcolor[HTML]{EFEFEF}  & \cellcolor[HTML]{EFEFEF}2.040 & \cellcolor[HTML]{EFEFEF}1.169 & \cellcolor[HTML]{EFEFEF}1.531 & \cellcolor[HTML]{EFEFEF}11.867 & \cellcolor[HTML]{EFEFEF}-5.04\% \\
\multirow{-2}{*}{\cellcolor[HTML]{EFEFEF}T14d}  & \cellcolor[HTML]{EFEFEF}0.189 & \cellcolor[HTML]{EFEFEF}0.128 & \cellcolor[HTML]{EFEFEF}0.155 & \cellcolor[HTML]{EFEFEF}0.569 & \cellcolor[HTML]{EFEFEF}-7.35\% \\ \hline
  & \textbf{1.965} & \textbf{1.110} & \textbf{1.394} & \textbf{11.141} & \textbf{-10.72\%} \\
\multirow{-2}{*}{T23}  & \textbf{0.182} & 0.121 & \textbf{0.140} & 0.572 & \textbf{-11.62}\% \\
\cellcolor[HTML]{EFEFEF} & \cellcolor[HTML]{EFEFEF}1.992 & \cellcolor[HTML]{EFEFEF}1.134 & \cellcolor[HTML]{EFEFEF}1.449 & \cellcolor[HTML]{EFEFEF}11.315 & \cellcolor[HTML]{EFEFEF}-8.70\% \\
\multirow{-2}{*}{\cellcolor[HTML]{EFEFEF}T23b}  & \cellcolor[HTML]{EFEFEF}0.184 & \cellcolor[HTML]{EFEFEF}0.125 & \cellcolor[HTML]{EFEFEF}\textbf{0.140} & \cellcolor[HTML]{EFEFEF}0.588 & \cellcolor[HTML]{EFEFEF}-10.08\% \\
  & 2.010 & 1.137 & 1.476 & 12.014 & -6.59\% \\
\multirow{-2}{*}{T23c} & 0.182 & \textbf{0.118} & 0.146 & 0.589 & -10.48\% \\
\cellcolor[HTML]{EFEFEF}  & \cellcolor[HTML]{EFEFEF}2.017 & \cellcolor[HTML]{EFEFEF}1.135 & \cellcolor[HTML]{EFEFEF}1.581 & \cellcolor[HTML]{EFEFEF}11.805 & \cellcolor[HTML]{EFEFEF}-5.32\% \\
\multirow{-2}{*}{\cellcolor[HTML]{EFEFEF}T23d} & \cellcolor[HTML]{EFEFEF}0.188 & \cellcolor[HTML]{EFEFEF}0.120 & \cellcolor[HTML]{EFEFEF}0.144 & \cellcolor[HTML]{EFEFEF}0.584 & \cellcolor[HTML]{EFEFEF}-9.94\% \\

\bottomrule
\hline
\end{tabular}
}
\end{footnotesize}
\end{center}
\end{minipage}

\end{table*}

We perform a strategic search on the trellis diagram for optimal stride configuration, as shown in Fig.~\ref{fig_Trellis_diagram} (a). All stride configurations are evaluated on the development set, and the results are reported in the left sub-table of Table~\ref{Table_all}. These results yield the following observations:

\textit{\textbf{Observation 1 --}} \textit{Models that prioritize preservation of temporal resolution over frequency resolution (indexed starting with `T') tend to outperform the default equal-stride configuration (indexed as MOD). Conversely, configurations that emphasize frequency resolution (indexed starting with `F') generally result in poorer performance.}
Fig.~\ref{fig_Trellis_diagram} (a) provides clear evidence that models utilizing stride configurations located in the upper-left partition of the trellis diagram prioritize the preservation of temporal resolution, resulting in a considerable advantage as indicated by the presence of a large bubble. In contrast, models positioned in the lower-right partition demonstrate an opposite trend. These observations strongly support the \textbf{\textit{Golden-Gemini Hypothesis}}, which posits that temporal resolution plays a more important role than frequency resolution in capturing the speaker characteristics of speech signals.

\textit{\textbf{Observation 2 --}} \textit{The performance of models with endpoints located on the boundary will significantly deteriorate.} 
As shown in Table~\ref{Table_all}, models indexed as T05, T15, T25, F52, F51, and F50 exhibit notable performance degradation compared to neighboring models on the trellis diagram. Unlike a TDNN that utilizes large channel numbers (e.g., 512, 1024, or 2048)~\cite{ECAPA_TDNN, MFA_TDNN, ECAPA_CNN_TDNN}, ResNet employs a smaller channel number (such as 32 or 64) at early stages for low-dimensional information representations~\cite{BUT2019system, depth_first, depth_first_conf, wespeaker}. This aligns with the design principle discussed in Section~\ref{Sec_mod_resnets} that emphasizes the importance of depth over width.
Consequently, when the temporal or frequency resolution is rapidly compressed, and constrained by a limited number of filters, it leads to the loss of information in that specific dimension. This results in a notable degradation of performance. 
Therefore, when designing a narrow ResNet with a smaller width, it is advisable to avoid the stride configurations located on the boundary.

\textit{\textbf{Observation 3 --}} \textit{Leveraging an optimal stride configuration effectively utilizes the computational resources of model size and FLOPs.} 
The trellis diagram in Fig~\ref{fig_Trellis_diagram} (a) clearly shows that models in the lower right region with large FLOPs and model size perform poorly. Even the largest model (indexed as F50), which employs an unfavorable stride configuration, performs the worst. On the contrary, models that preserve temporal resolution achieve good performance with less complexity than the baseline.

\textit{\textbf{Observation 4 --}} \textit{Models indexed as T14 and T23 show the best performance among all the models.} 
This supports the \textbf{\textit{Golden-Gemini Hypothesis}} that there exist operational states that yield optimal performance for ASV. We refer to these two endpoints on the trellis diagram representing a pair of optimal operational states as the \textit{Golden Gemini}. 

Points closer to the start points are not explored for two reasons. Firstly, T13 shows inferior performance compared to \textit{Golden Gemini}. Secondly, points closer to the start points would significantly increase computational complexity, resulting in decreased efficiency compared to utilizing a deeper model with the proposed \textit{Golden-Gemini} stride configuration.

In the right sub-table of Table~\ref{Table_all}, the testing results for all stride configurations are presented. It is evident that across all four testing sets, covering in-domain, out-domain, large-scale, and hard-case scenarios, the testing results exhibit a consistent trend similar to that observed in the development set. This consistency strongly supports the above observations.

\subsection{Evaluation on Different Paths towards Golden Gemini}
\label{subsec_eval_golden_gemini}

There are multiple paths leading to the \textit{Golden Gemini}, as shown in Fig.~\ref{fig_Trellis_diagram} (b), each representing a stride configuration. As discussed in Section~\ref{subsec_Trellis_Diagrams}, we further investigate these paths to assess the impact of early or late downsampling. The experimental results of the development set are presented in the left sub-table of Table~\ref{table_Golden_Gemini}. Following are the two observations:

\textit{\textbf{Observation 5 --}} \textit{All paths towards the \textit{Golden Gemini} points outperform the baseline (indexed as MOD) that uses an equal-stride configuration.}
This observation supports \textbf{\textit{Golden-Gemini Hypothesis}} that the pair of operational states engage in competition and yield optimal performance.

\textit{\textbf{Observation 6 --}} \textit{Different path options offer the flexibility to trade off between FLOPs and performance, with increased FLOPs generally resulting in improved results.} This flexibility in model design allows for better adaptation to specific application scenarios. In addition, previous work demonstrates superiority by comparing FLOPs as a metric~\cite{resnet, MFA_TDNN, depth_first, depth_first_conf, wang2023cam++, liu2022convnet, res2net, swin_Transformer, 9746247}. This practice is based on the common understanding that bigger FLOPs often correlate with better performance. However, rather than simply increasing FLOPs, experimental results show that an optimal stride configuration utilizes FLOPs more efficiently.

The testing results reported in the right sub-table of Table~\ref{table_Golden_Gemini} demonstrate a consistent trend similar to that observed on the development set. This further verifies the two observations mentioned above.
In addition, all the stride configurations depicted in both trellis diagrams in Fig.~\ref{fig_Trellis_diagram} are visualized in Fig.~\ref{fig_compare_all_stride}, comparing their performance, model size, and FLOPs. 
Among these configurations, the \textit{Golden-Gemini} T14c achieves average EER/minDCF reductions of 5.78\%/14.37\% over the modified ResNet baseline (indexed as MOD) across all four test sets while reducing the model size by 9.8\% and the computational complexity by 4.2\%. 
Considering the efficacy and efficiency, we designate the T14c stride configuration as the principal stride configuration in this work.
Networks that adopt the proposed \textit{Golden-Gemini} stride configurations are referred to as the \textit{Gemini} networks, such as the \textbf{\textit{Gemini} ResNet}.
The structure comparison of the proposed \textit{Gemini} ResNet with T14c stride configuration, original ResNet~\cite{resnet}, and modified ResNet~\cite{BUT2019system, wespeaker} is presented in Table~\ref{table_structure}.

\begin{figure*}[t]
\centerline{\includegraphics[scale=0.047]{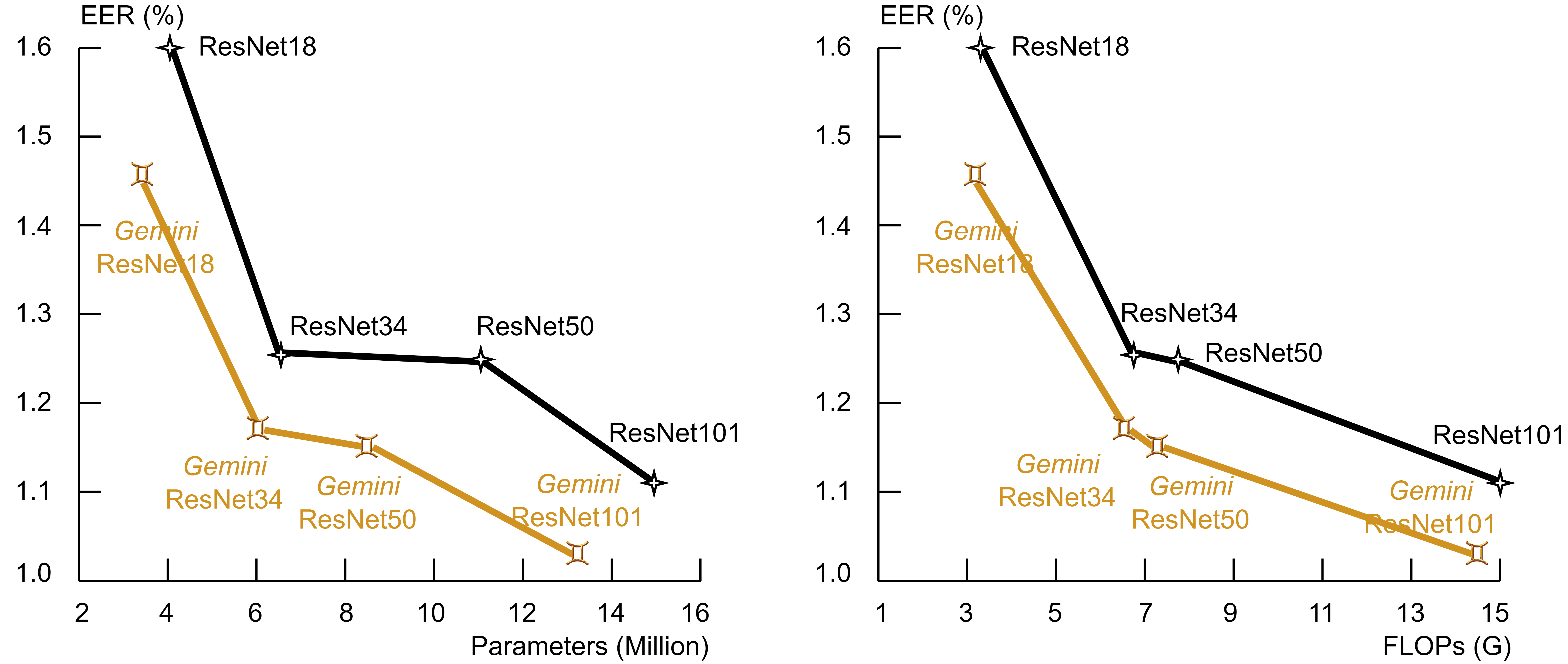}}
\vspace{-2mm}
\caption{Performance and complexity comparison of proposed \textit{Gemini} ResNet and modified ResNet\cite{BUT2019system, wespeaker} with different model sizes on Vox1-E test set.}
\label{fig_diff_size}
\vspace{-3mm}
\end{figure*}

\begin{table*}[h]
\caption{Performance in EER(\%) and minDCF of different sizes of ResNet models with equal-stride configuration or the proposed \textit{Golden-Gemini} stride configuration (T14c) on VoxCeleb1, SITW and CNCeleb test sets. Avg. Reduction indicates averaged relative reduction across all five sets over the benchmark.}
\label{table_diff_size}
\centering
\begin{footnotesize}
\renewcommand\arraystretch{1.08}
\setlength{\tabcolsep}{0.85mm}{
\begin{tabular}{clccccccc}
\hline
\toprule
\multirow{2}{*}{Model} & \multirow{2}{*}{\begin{tabular}[c]{@{}l@{}}Params\\ (Million)\end{tabular}} & FLOPs (G) & Vox1-O & Vox1-H & Vox1-E & SITW & CNCeleb & Avg. Reduction \\ \cline{3-9} 
 &  & 2s/3s & EER/minDCF & EER/minDCF & EER/minDCF & EER/minDCF & EER/minDCF & EER/minDCF  \\ \hline

\midrule
 
ResNet18 & 4.11 & 2.22/3.30 & 1.760/0.177 & 2.785/0.244 & 1.600/0.170 & 2.132/0.210 & 12.301/0.657 & Benchmark \\
\textbf{\textit{Gemini} ResNet18} & \textbf{3.45 (-16.1\%)} & \textbf{2.17/3.25} & \textbf{1.319/0.139} & \textbf{2.474/0.226} & \textbf{1.462/0.153} & \textbf{2.050/0.190} & \textbf{12.211/0.592} & -9.89\%/-11.57\% \\ \hline
ResNet50 & 11.13 & 5.22/7.76 & 1.329/0.141 & 2.213/0.205 & 1.249/0.134 & 1.613/0.158 & 11.856/0.692 & Benchmark \\
\textbf{\textit{Gemini} ResNet50} & \textbf{8.51 (-23.5\%)} & \textbf{4.92/7.35} & \textbf{1.196/0.121} & \textbf{2.016/0.189} & \textbf{1.147/0.119} & \textbf{1.449/0.145} & \textbf{11.608/0.603} & -7.87\%/-11.22\% \\ \hline
ResNet101 & 15.89 & 10.07/15.00 & 1.101/0.100 & 2.051/0.194 & 1.121/0.121 & 1.367/0.140 & 11.884/0.633 & Benchmark \\
\textbf{\textit{Gemini} ResNet101} & \textbf{13.27 (-16.5\%)} & \textbf{\ 9.72/14.54} & \textbf{0.962/0.099} & \textbf{1.836/0.167} & \textbf{1.035/0.108} & \textbf{1.320/0.125} & \textbf{11.625/0.553} & -7.26\%/-9.88\% \\ 

\bottomrule
\hline
\end{tabular}
}
\end{footnotesize}
\vspace{-1mm}
\end{table*}

\vspace{-2mm}
\begin{table*}[t]
\centering
\caption{Performance in EER(\%) and minDCF of the networks with equal-stride configuration and proposed \textit{Golden-Gemini} T14c stride configuration under different conditions on VoxCeleb1, SITW, and CNCeleb test sets. Aug. indicates whether the system is trained with data augmentations. Avg. Reduction means the average relative reduction across all five sets over the benchmark.}
\label{table_compatibility}
\begin{footnotesize}
\renewcommand\arraystretch{1.08}
\setlength{\tabcolsep}{0.75mm}{
\begin{tabular}{cccccccccc}
\hline
\toprule
\multirow{2}{*}{Model} & \multirow{2}{*}{Aug.} & \multirow{2}{*}{\begin{tabular}[c]{@{}c@{}}Params\\ (Million)\end{tabular}} & FLOPs (G) & Vox1-O & Vox1-H & Vox1-E & SITW & CNCeleb & Avg. Reduction \\ \cline{4-10} 
 &  &  & 2s/3s & EER/minDCF & EER/minDCF & EER/minDCF & EER/minDCF & EER/minDCF & EER/minDCF \\ \hline
\midrule
ResNet34~\cite{BUT2019system} & \textbf{$\times$} & 6.63 & 4.63/6.88 & 1.489/0.155 & 2.500/0.224 & 1.423/0.158 & 2.378/0.208 & 12.737/0.639 & Benchmark \\
\textbf{\textit{Gemini} ResNet34} & \textbf{$\times$} & \textbf{5.98} & \textbf{4.41/6.59} & \textbf{1.375/0.132} & \textbf{2.261/0.209} & \textbf{1.294/0.139} & \textbf{2.102/0.189} & \textbf{12.271/0.628} & -8.30\%/-8.88\% \\ \hline
ResNet34 + SE\cite{SEnet} & $\checkmark$ & 6.79 & 4.63/6.89 & 1.287/0.141 & 2.512/0.241 & 1.370/0.157 & 1.640/0.187 & \textbf{12.408/0.688} & Benchmark \\
\textbf{\textit{Gemini} ResNet34 + SE\cite{SEnet}} & $\checkmark$ & \textbf{6.14} & \textbf{4.41/6.60} & \textbf{1.053/0.104} & \textbf{2.264/0.219} & \textbf{1.223/0.143} & \textbf{1.531/0.165} & 13.078/0.703 & -8.00\%/-10.75\% \\ \hline
Res2Net34\cite{res2net} & $\checkmark$ & 6.57 & 4.74/7.05 & 1.071/0.103 & 2.073/0.195 & 1.184/0.129 & 1.524/0.157 & 11.571/\textbf{0.568} & Benchmark \\
\textbf{\textit{Gemini} Res2Net34} & $\checkmark$ & \textbf{5.92} & \textbf{4.46/6.68} & \textbf{1.048/0.100} & \textbf{1.914/0.176} & \textbf{1.092/0.115} & \textbf{1.422/0.135} & \textbf{11.135}/0.577 & -5.60\%/-7.18\% \\ \hline
ResNet34 + xi\cite{xivector} & $\checkmark$ & 7.30 & 4.66/6.93 & 1.159/0.111 & 2.192/0.215 & 1.288/0.136 & 1.602/0.161 & 12.627/0.652 & Benchmark \\
\textbf{\textit{Gemini} ResNet34 + xi\cite{xivector}} & $\checkmark$ & \textbf{6.31} & \textbf{4.47/6.69} & \textbf{1.101/0.105} & \textbf{2.100/0.199} & \textbf{1.169/0.124} & \textbf{1.480/0.149} & \textbf{11.902/0.644} & -6.37\%/-6.07\% \\  \hline

SD-ResNet38\cite{liu2022convnet} & $\checkmark$ & 7.37 & 5.20/7.74 & 1.202/0.130 & 2.133/0.203 & 1.187/0.131 & 1.586/0.161 & 11.580/0.618 & Benchmark \\
\textbf{\textit{Gemini} SD-ResNet38\cite{liu2022convnet}} & $\checkmark$ & \textbf{6.72} & \textbf{4.97/7.43} & \textbf{1.085/0.099} & \textbf{1.974/0.185} & \textbf{1.130/0.117} & \textbf{1.523/0.147} & \textbf{11.507/0.553} & -5.32\%/-12.60\% \\

\bottomrule
\hline
\end{tabular}
}
\end{footnotesize}
\vspace{-2mm}
\end{table*}

\subsection{Evaluation on Compatibility}
{The changes in time and frequency resolutions occur once per stage in both ResNet and its variant networks, such as DF-ResNet~\cite{depth_first}, Res2Net~\cite{res2net}, and SD-ResNet~\cite{liu2022convnet}. Given that the \textit{Golden Gemini}  is concluded from investigating the significance of time and frequency resolution for speaker verification, it is expected to apply to all ResNet series networks that still adhere to the original ResNet's five-stage structural design.}
This subsection aims to confirm the consistently superior performance of the proposed \textit{Golden-Gemini} stride configuration across various conditions and its compatibility with different techniques. 
{We exemplify with the \textit{Golden Gemini} T14c stride configuration, conducting experiments to compare} the models using the \textit{Golden Gemini} T14c and the default equal-stride configuration under the following conditions:

\textbf{Different model sizes}. 
The ResNet models have different depths, resulting in different model sizes and computational resource requirements. For any proposed new method, adapting to ResNet models of various sizes is important as it allows for trade-offs between performance and complexity, enabling better adaptation to different application scenarios. We further extend the application of the proposed \textit{Golden-Gemini} stride configuration from ResNet34 to a smaller model (ResNet18) and the larger models (ResNet50 and ResNet101). The experimental results are presented in Table~\ref{table_diff_size} and Fig.~\ref{fig_diff_size}. The results demonstrate that the proposed \textit{Golden-Gemini} stride configuration consistently improves the performance by an average of 7.70\%/11.76\% EER/minDCF reduction across the entire range of model sizes, while reducing parameters and FLOPs by 16.5\% and 4.1\%, respectively.

\textbf{Data augmentations}.
\label{subsubsec_aug}
The training of neural networks benefits from data augmentations~\cite{SpecAugment}. All previous experiments are trained with augmented data as described in Section~\ref{subsec_training_strategy}. We conduct training without data augmentation to assess the compatibility of \textit{Golden Gemini}, and the results are shown in Table~\ref{table_compatibility}. It is observed that the \textit{Golden-Gemini} stride configuration achieves an average relative reduction of 8.30\%/.88\% in EER/minDCF across five sets, and reduces complexity.

\textbf{Squeeze-and-excitation (SE) attention module~\cite{SEnet}}. 
SE is one of the most widely used attention modules. We validate the compatibility of the proposed \textit{Golden-Gemini} stride configuration with SE (reduction ratio $r=4$), and the results are shown in Table~\ref{table_compatibility}. We can observe that the proposed \textit{Golden Gemini} outperforms the equal-stride configuration on most of the test sets, with an average EER/minDCF reduction of 8.00\%/10.75\%. However, the SE block does not improve the performance, which may require further investigation.

\textbf{A different backbone network -- Res2Net~\cite{res2net}}.
The proposed \textit{Golden-Gemini} stride configuration is not limited to ResNet models and can be applied to other 2D CNN-based models as well. Res2Net~\cite{res2net} is a well-known 2D CNN-based architecture recognized for its ability to extract multi-scale features. In the design of multi-scale frequency-channel attention TDNN (MFA-TDNN)~\cite{MFA_TDNN} and multi-scale feature aggregation convolution-augmented transformer (MFA-Conformer)~\cite{zhang22h_interspeech}, multi-scale features have been proven to benefit ASV. Previous studies have explored the application of Res2Net in ASV~\cite{resnext_n_res2net_ASV, 9413832, chen2023enhanced}. We compare the Res2Net34 model using the default equal-stride configuration with that using the proposed \textit{Golden-Gemini} stride configuration. The scale ($s$) of Res2Net is set to 4. The results in Table~\ref{table_compatibility} show that the \textit{Golden-Gemini} stride configuration improves performance while reducing complexity compared to the equal-stride configuration. Additionally, Res2Net34 shows better performance in ASV compared to ResNet34.

\textbf{A different temporal aggregation layer -- xi posterior inference (xi)~\cite{xivector}}. 
As introduced in Section~\ref{sec_Introduction}, an embedding extractor network consists of three components -- an encoder, a temporal aggregation layer, and a decoder. The previous experiments focus on the encoder component, and for a fair comparison, a default temporal statistics pooling~\cite{temporal_statistics_pooling} is applied across all experiments. We further validate the compatibility of the proposed \textit{Golden Gemini} with another temporal aggregation method -- xi posterior inference, which is designed to estimate uncertainty~\cite{xivector}. 
The experimental results shown in Table~\ref{table_compatibility} demonstrate the consistently superior performance of \textit{Golden Gemini} over the equal-stride configuration while reducing the model size by 13.6\% and FLOPs by 4.1\%.

\textbf{A micro design -- separate downsampling (SD)~\cite{liu2022convnet}}.
Unlike ResNet~\cite{resnet}, which performs downsampling at the first 2D CNN layer in each stage, Swin Transformer~\cite{swin_Transformer} introduces a separate downsampling layer between stages. This micro design is also extended to ResNet, resulting in notable improvements~\cite{liu2022convnet}. In this work, we explore this micro design for ASV by implementing four $3\times3$ 2D CNN layers between the five stages and name SD-ResNet. It is worth noting that this modification adds four additional 2D CNN layers, resulting in the expansion of the ResNet\textbf{34}~\cite{BUT2019system, wespeaker} architecture to SD-ResNet\textbf{38}. The results in Table~\ref{table_compatibility} demonstrate that SD-ResNet outperforms the modified ResNet~\cite{wespeaker} (indexed as MOD in Table~\ref{Table_all}). Moreover, the integration of \textit{Golden Gemini} leads to additional improvements in terms of EER/minDCF, with averaged reductions of 5.32\% and 12.60\%, respectively.

In summary, the experimental results validate the compatibility of the proposed \textit{Golden-Gemini} stride configuration with various existing techniques and training conditions.
\textit{Golden Gemini} consistently improves performance while reducing complexity.
Its superiority can be attributed to the importance of temporal resolution. By maintaining temporal resolution, \textit{Golden Gemini} ensures adequate representations of both vocal tract features and learned speaker characteristics across various scales of local time regions and is expected to further benefit the temporal aggregation layer, leading to significant performance improvements.

\subsection{New SOTA Benchmark}
\label{subsec_new_SOTA}
DF-ResNet~\cite{depth_first, depth_first_conf} is a series of powerful SOTA models introduced in Section~\ref{Sec_mod_resnets}. We first re-implement a small venison, namely DF-ResNet59. The results reported in Table~\ref{table_dfresnet} demonstrate that
the re-implemented DF-ResNet model slightly outperforms the one reported in~\cite{depth_first,depth_first_conf}. Notably, we do not apply SpecAugment~\cite{SpecAugment} which is used in~\cite{depth_first,depth_first_conf}. SpecAugment has been proven effective in automatic speech recognition (ASR)~\cite{SpecAugment}. However, it can have adverse effects on the fundamental frequency of the audio, which is a critical characteristic for speaker discrimination~\cite{ASV_SUBTOOLS}. Prior work~\cite{wespeaker} shows that combining SpecAugment with other augmentation methods in ASV can pose compatibility challenges. Our experiments demonstrate a similar trend.

\begin{table}[t]

\caption{The structure comparison between DF-ResNet~\cite{depth_first} with default equal-stride configuration and the proposed \textit{Golden-Gemini} stride configuration. SD and dw indicate separate downsampling and depth-wise convolution, respectively~\cite{depth_first}.}
\label{table_structure_dfresnet}
\vspace{-4.2 mm}
\begin{center}
\begin{scriptsize}
\setlength{\tabcolsep}{0.7mm}{
\renewcommand{\arraystretch}{1.06}{
\begin{tabular}{cc|cc|cc}
\hline
\toprule
\multirow{2}{*}{Stage} & \multirow{2}{*}{Layer} & \multicolumn{2}{c}{DF-ResNet182\tablefootnote{\label{fnote_layers} For the re-implemented DF-ResNet and proposed \textit{Gemini} DF-ResNet models, we count the separate downsampling layers as part of the total layer count. 
This differs from the counting method used in the original DF-ResNet~\cite{depth_first, depth_first_conf}. As an example, the DF-ResNet\textbf{179} in~\cite{depth_first, depth_first_conf} is referred to as DF-ResNet\textbf{182} in this work. However, for the experimental results cited in Table~\ref{table_dfresnet}, we follow the original work~\cite{depth_first, depth_first_conf}.}
} & \multicolumn{2}{c}{\textit{Gemini} DF-ResNet183\footref{fnote_layers}} \\
 & & Stride & Output   & Stride & Output\\ 
\hline
\midrule
conv1 & 3$\times$3, 32 & (1,1) & 32$\times$F$\times$T & (1,1)& 32$\times$F$\times$T \\ \hline

 & 3$\times$3, 32 (SD) &- & -&(2,1)& 32$\times$F/2$\times$T \\ 
conv2 & $\begin{bmatrix}1\times1,   128\\3\times3, 128 (dw)  \\1\times1, 32  \end{bmatrix}$ $\times$3 &(1,1)& 32$\times$F$\times$T     &(1,1)& 32$\times$F/2$\times$T \\ \hline

 & 3$\times$3, 64 (SD) &(2,2)& 64$\times$F/2$\times$T/2 &(2,2)& 64$\times$F/4$\times$T/2 \\ 
conv3 & $\begin{bmatrix}1\times1,   256\\3\times3, 256 (dw)  \\1\times1, 64  \end{bmatrix}$ $\times$8 &(1,1)& 64$\times$F/2$\times$T/2  &(1,1)& 64$\times$F/4$\times$T/2 \\ \hline

 & 3$\times$3, 128 (SD) &(2,2)& 128$\times$F/2$\times$T/2 &(2,1)& 128$\times$F/4$\times$T/2 \\ 
conv4 & $\begin{bmatrix}1\times1,   512\\3\times3, 512 (dw)  \\1\times1, 128 \end{bmatrix}$ $\times$45 &(1,1)& 128$\times$F/4$\times$T/4 &(1,1)&128$\times$F/8$\times$T/2 \\ \hline

 & 3$\times$3, 256 (SD) &(2,2)& 256$\times$F/2$\times$T/2 &(2,1)& 256$\times$F/4$\times$T/2 \\ 
conv5 & $\begin{bmatrix}1\times1,  1024\\3\times3, 1024 (dw) \\1\times1, 256 \end{bmatrix}$ $\times$3 &(1,1)& 256$\times$F/8$\times$T/8 &(1,1)&256$\times$F/16$\times$T/2 \\ \hline

\multicolumn{2}{c|}{Temporal Statistics Pooling Layer}& N/A & 256$\times$F/8$\times$2 & N/A & 256$\times$F/16$\times$2\\\hline
\multicolumn{2}{c|}{Fully Connected Layer}  & \multicolumn{2}{c|}{(5120, 256)} & \multicolumn{2}{c}{ (2560, 256)}\\\hline
\multicolumn{2}{c|}{\# Parameters} &\multicolumn{2}{c|}{9.84$\times 10^6$ } & \multicolumn{2}{c}{\textbf{9.20}$\times 10^6$ }\\
\multicolumn{2}{c|}{FLOPs (2s / 3s)} & \multicolumn{2}{c|}{8.64$\times 10^9$ / 12.87$\times 10^9$ } & \multicolumn{2}{c}{\textbf{8.25}$\times 10^9$ / \textbf{12.34}$\times 10^9$ }\\
\bottomrule
\hline
\end{tabular}
}
}
\end{scriptsize}
\end{center}
\vspace{-4 mm}
\end{table}

\begin{table}[t]
\centering
\caption{Performance in EER(\%) and minDCF of the proposed \textit{Golden Gemini} DF-ResNet and SOTA systems on VoxCeleb1 test sets. Models with our proposed \textit{Golden-Gemini} stride configuration are highlighted in \colorbox{gray!30}{grey}
~\tablefootnote{\label{fnote_pretrain} Pre-trained models and codes of the proposed \textit{Gemini} DF-ResNet are available at https://github.com/Tianchi-Liu9/Golden-Gemini-for-Speaker-Verification and https://github.com/wenet-e2e/wespeaker.}.
}
\vspace{-0 mm}
\label{table_dfresnet}
\begin{scriptsize}
\renewcommand\arraystretch{1.06}
\setlength{\tabcolsep}{0.5mm}{
\begin{tabular}{cclclclc}
\cline{1-6}
\hline
\toprule

\multirow{2}{*}{System} & \multirow{2}{*}{Para.} & \multicolumn{2}{c}{Vox1-O} & \multicolumn{2}{c}{Vox1-E} & \multicolumn{2}{c}{Vox1-H} \\ \cline{3-8} 
 &  & EER & minDCF & EER & minDCF & EER & minDCF \\ 
 
\hline
\midrule

Res2Net-14w8s~\cite{resnext_n_res2net_ASV} & 5.6 & 1.60 & 0.178 & 1.60 & 0.184 & 2.83 & 0.280 \\
ResNet18~\cite{depth_first_conf} & 4.11 & 1.48 & 0.174 & 1.52 & 0.175 & 2.72 & 0.244 \\
ECAPA-TDNN (C=512)~\cite{ECAPA_TDNN} & 6.2 & 1.01 & 0.127 & 1.24 & 0.142 & 2.32 & 0.218 \\
MFA-TDNN (Lite)~\cite{MFA_TDNN} & 5.93 & 0.968 & 0.091 & 1.138 & 0.121 & 2.174 & 0.199 \\ \hline
DF-ResNet56\footref{fnote_layers}~\cite{depth_first, depth_first_conf} & 4.49 & 0.96 & 0.103 & 1.09 & 0.122 & 1.99 & 0.184 \\
DF-ResNet59\footref{fnote_layers} (re-implemented) & 4.69 & 0.973 & 0.097 & 1.060 & 0.120 & 1.866 & 0.175 \\
\rowcolor[HTML]{EFEFEF}\textbf{\textit{Gemini} DF-ResNet60}\footref{fnote_layers} & \textbf{4.05} & \textbf{0.941} & \textbf{0.089} & \textbf{1.051} &\textbf{0.116} & \textbf{1.799} & \textbf{0.166} \\ 
\hline
\hline
E-TDNN~\cite{ECAPA_TDNN} & 6.8 & 1.49 & 0.160 & 1.61 & 0.171 & 2.69 & 0.242 \\
Res2Net-26w8s~\cite{resnext_n_res2net_ASV} & 9.3 & 1.45 & 0.147 & 1.47 & 0.169 & 2.72 & 0.272 \\
ResNet34~\cite{depth_first_conf} & 6.63 & 0.96 & 0.089 & 1.01 & 0.121 & 1.86 & 0.177 \\
H/ASP AP+softmax~\cite{9413948} & 8.0 & 0.88 & - & 1.07 & - & 2.21 & - \\
MFA-TDNN (Standard)~\cite{MFA_TDNN} & 7.32 & 0.856 & 0.092 & 1.083 & 0.118 & 2.049 & 0.190 \\
RecXi with $\mathcal{L}_{\rm ssp}$~\cite{liu2023disentangling} & 7.06 & 0.984 & 0.091 & 1.075 & 0.114 & 1.857 & 0.179 \\
PCF-ECAPA (C=512)~\cite{10095051} & 8.9 & 0.718 & 0.086 & 0.792 & 0.114 & 1.802 & 0.175 \\ 
CAM++~\cite{wang2023cam++} & 7.18 & 0.73 & 0.091 & 0.89 & 0.100 & 1.76 & 0.173\\
\hline
DF-ResNet110\footref{fnote_layers}~\cite{depth_first, depth_first_conf} & 6.98 & 0.75 & 0.070 & 0.88 & 0.100 & 1.64 & 0.156 \\
\rowcolor[HTML]{EFEFEF}\textbf{\textit{Gemini} DF-ResNet114}\footref{fnote_layers}& \textbf{6.53} & \textbf{0.686} & \textbf{0.067} & \textbf{0.863} & \textbf{0.097} & \textbf{1.490} & \textbf{0.144} \\ 
\hline
\hline
E-TDNN (large)~\cite{ECAPA_TDNN} & 20.4 & 1.26 & 0.140 & 1.37 & 0.149 & 2.35 & 0.215 \\
ResNet18~\cite{ECAPA_TDNN} & 13.8 & 1.47 & 0.177 & 1.60 & 0.179 & 2.88 & 0.267 \\
ResNet34~\cite{ECAPA_TDNN} & 23.9 & 1.19 & 0.159 & 1.33 & 0.156 & 2.46 & 0.229 \\
ECAPA-TDNN (C=1024)~\cite{ECAPA_TDNN} & 14.7 & 0.87 & 0.107 & 1.12 & 0.132 & 2.12 & 0.210 \\
MFA-Conformer (1/2)~\cite{zhang22h_interspeech} & 20.5 & 0.64 & 0.081 & 1.29 & 0.137 & 1.63 & 0.153 \\
P-vectors (SFA)~\cite{pvector2023intspch} &15.1 & 0.856 & 0.120 & 1.117 &0.120 &2.112 & 0.208 \\
ResNet52-C2D-32~\cite{10095415} &10.34 & 0.771 & 0.107 & 0.939 & 0.111 &1.816 & 0.180 \\
SKA-TDNN~\cite{10023305} & 34.9 & 0.78 & - & 0.90 & - & 1.74 & - \\
SimAM-ResNet34 (GSP)~\cite{9746294} & 21.54 & 0.718 & 0.071 & 0.993 & 0.103 & 1.647 & 0.159 \\
DS-TDNN-L~\cite{li2023dstdnn} & 20.5 & 0.64 & 0.082 & 0.93 & 0.112 & 1.55 & 0.149 \\
PCF-ECAPA (C=1024)~\cite{10095051} & 22.2 & 0.718 & 0.089 & 0.891 & 0.102 & 1.707 & 0.175 \\
{NEMO}~\cite{10096659} & 15.88 & 0.74 & 0.110 & 0.90 & 0.105 & 1.90 & 0.189 \\ 
{Branch-ECAPA-TDNN(b)}~\cite{yao23_interspeech} & 24.11 & 0.72 & 0.084 & 0.92 & 0.098 & 1.69 & 0.166 \\ 
{ECAPA++ (Big)}~\cite{Ecapa++} & 23.9 & 0.65 & 0.080 & 0.84 & 0.098 & 1.54 & 0.154 \\ 
\hline
DF-ResNet179\footref{fnote_layers}~\cite{depth_first, depth_first_conf} & 9.84 & 0.62 & \textbf{0.061} & \textbf{0.80 }& \textbf{0.090} & 1.51 & 0.148 \\
\rowcolor[HTML]{EFEFEF}\textbf{\textit{Gemini} DF-ResNet183}\footref{fnote_layers} & \textbf{9.20} & \textbf{0.596} & 0.065 & 0.806 & \textbf{0.090} & \textbf{1.440} & \textbf{0.137} \\ 

\bottomrule
\hline
\end{tabular}
}
\end{scriptsize}
\vspace{-4 mm}
\end{table}

Similar to other ResNet models, DF-ResNet adopts the default equal-stride configuration, treating temporal and frequency dimensions equally. {By replacing the stride configuration  with our proposed \textit{Golden-Gemini} T14c, we see a notable 4.9\% average performance boost and a 7.6\% reduction in model size, as detailed in Table~\ref{table_dfresnet}. Also, Table~\ref{table_structure_dfresnet} shows a 4.5\% decrease in FLOPs. It's important to note that DF-ResNet, our chosen baseline, achieves SOTA performance with a relatively small model size, emphasizing its meticulous design and efficiency. In this context, achieving further performance gains becomes challenging given the already very low EER and minDCF. Further analysis indicates that, as the model size increases from the smallest to largest, relative performance improvements decrease from 7.4\% to 5.8\%, and then to 1.7\%. This trend aligns with the inherent difficulty of achieving significant performance improvements over a robust baseline and low EER/minDCF. Nevertheless, our proposed \textit{Golden-Gemini} stride configuration still brings improvements, securing the best performance among all systems.}
This outcome supports the remarkable capabilities of the \textit{Golden-Gemini} stride configuration and \textbf{\textit{Golden-Gemini Hypothesis}}, which emphasizes the critical significance of temporal resolution in attaining superior results in ASV.

\subsection{\textit{Golden-Gemini} Guiding Principles}

The experiments conducted above consistently demonstrate the superiority of the proposed \textit{Golden Gemini} over the default equal-stride configuration from various perspectives.
The underlying logic behind the \textit{Golden Gemini} is the utilization of a series of guiding principles that align with the natural properties of speech signals for designing 2D CNN-based networks for ASV. Based on the aforementioned observations, we summarize the \textit{Golden-Gemini} guiding principles as follows:
\begin{itemize}
\item Preserve sufficient temporal resolutions during the feature representation instead of preserving the frequency resolution.
\item Avoid frequently diminishing any dimension at the early stage when using a narrow network.
\item A correct stride configuration surpasses mere FLOP increments. Prioritize the adoption of the optimal stride configuration followed by the trade-off between FLOPs and performance according to computation resources.
\end{itemize}

\section{Conclusion}
We investigate efficient stride configurations for speaker verification. Through a strategic search on a trellis diagram, we analyze the impact of temporal and frequency resolution on the ASV performance. Experimental results on the VoxCeleb, SITW, and CNCeleb test sets highlight the significance of the temporal resolution. This leads us to identify two points, named \textit{Golden Gemini}, representing two series of optimal stride configurations for ASV.
We also present a set of guiding principles that comprehensively describe the \textit{Golden Gemini} for designing 2D ResNet for ASV.
Further experiments demonstrate the consistent superiority and excellent compatibility of the proposed \textit{Golden Gemini} with various structures across different conditions.  Moreover, our approach is simple yet effective and can be easily applied to any 2D ResNet architecture style, offering improved performance while reducing model complexity. Based on the \textit{Golden-Gemini} guiding principles, we introduce a powerful benchmark for ASV, namely the \textit{Gemini} DF-ResNet. These findings indicate the promising value of our method in real-world applications. {Additionally, the significance of time and frequency resolutions may extend beyond speaker verification, holding great potential for related tasks such as speaker diarization, speaker extraction, emotion recognition, and speech anti-spoofing.}

\bibliographystyle{IEEEtran}
\bibliography{main}











\vspace{-15 pt}

\begin{IEEEbiography}[{\includegraphics[width=1in,height=1.25in,clip,keepaspectratio]{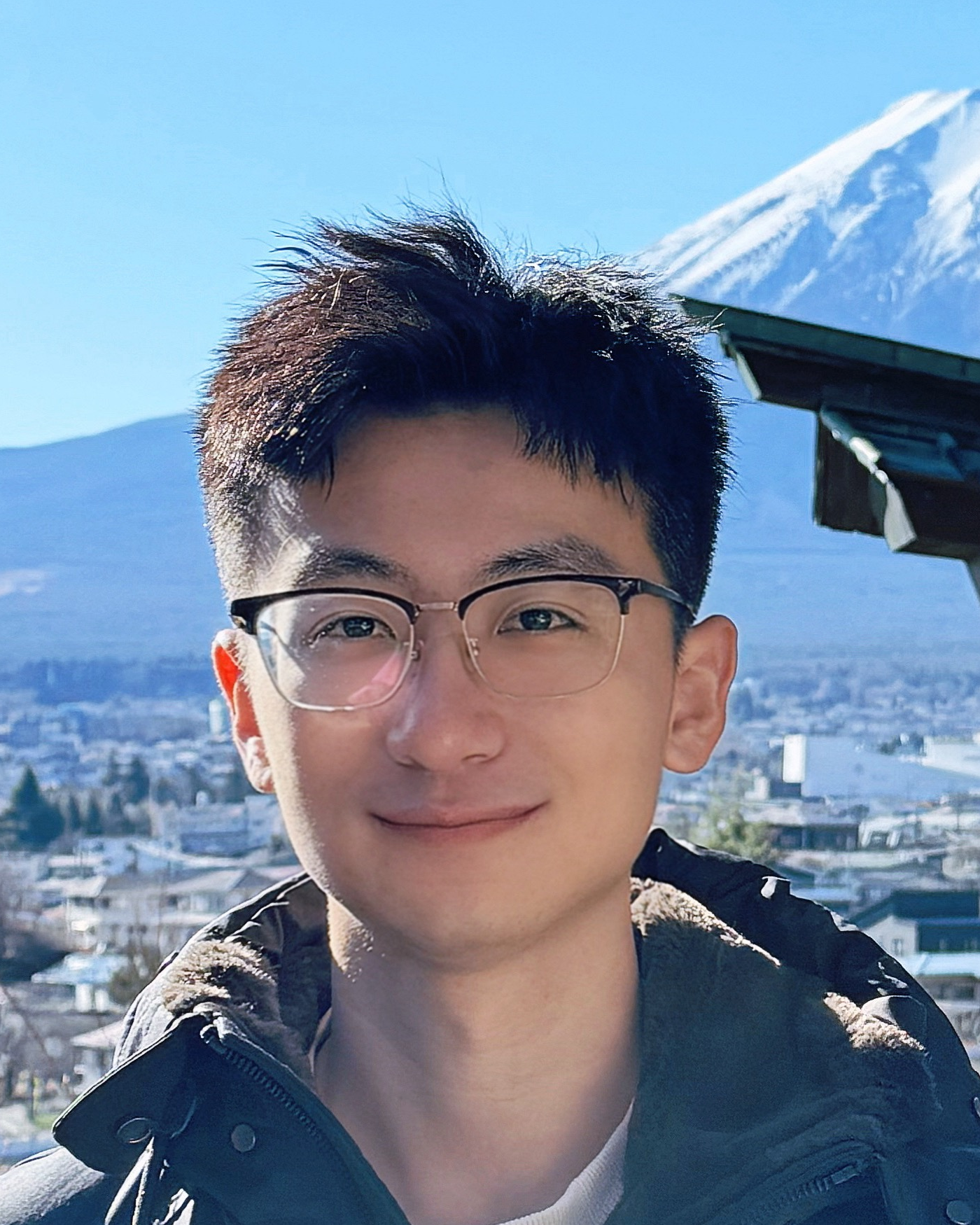}}]{Tianchi Liu} (Student Member, IEEE) received the M.Sc. degree in 2019 from the National University of Singapore, Singapore. He is currently working toward the Ph.D. degree with the Department of Electrical and Computer Engineering, National University of Singapore, Singapore. He is also a Senior Research Engineer at Institute for Infocomm Research (I$^2$R), Agency for Science, Technology, and Research (A$^\star$STAR), Singapore. 
His research interests include speaker recognition, speech anti-spoofing, audio-visual representation learning, large language model (LLM), and speech foundation model.
\end{IEEEbiography}


\begin{IEEEbiography}[{\includegraphics[width=1in,height=1.25in,clip,keepaspectratio]{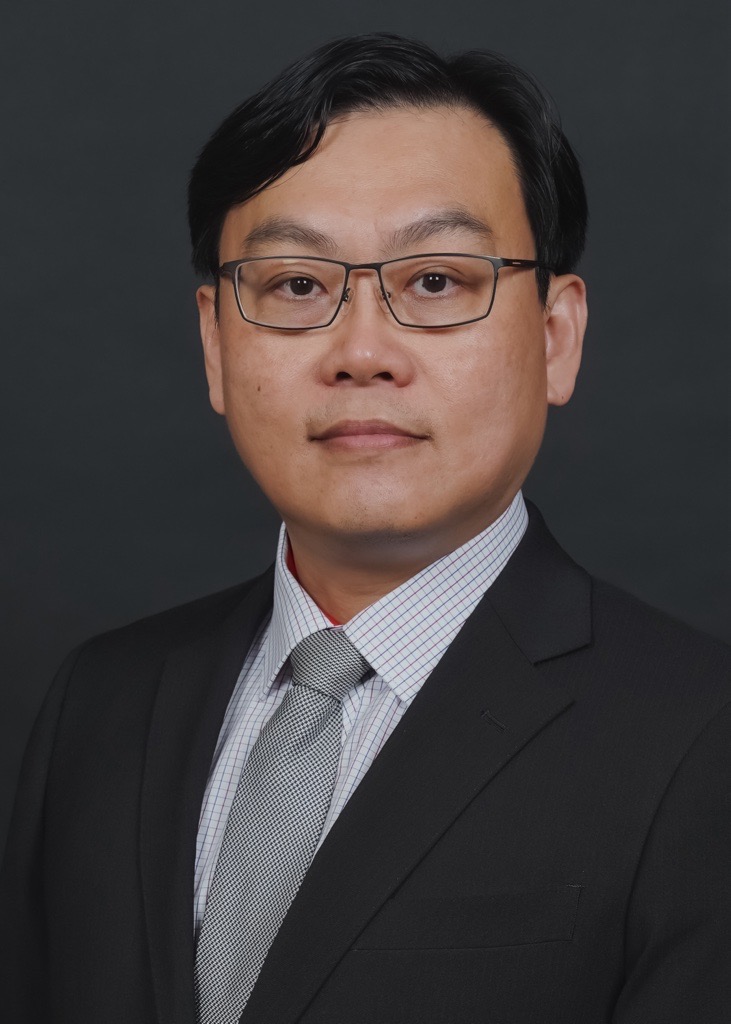}}]{Kong Aik Lee}
(Senior Member, IEEE) received his Ph.D. from Nanyang Technological University, Singapore, in 2006. From 2006 to 2018, he was a Research Scientist and then a Strategic Planning Manager (concurrent appointment) with the Institute for Infocomm Research, Singapore. From 2018 to 2020, he was a Senior Principal Researcher at the Data Science Research Laboratories, NEC Corporation, Tokyo, Japan. He is currently an Associate Professor at the Hong Kong Polytechnic University, Hong Kong. Before joining PolyU, he was an Associate Professor at the Singapore Institute of Technology, Singapore, while holding a concurrent appointment as a Principal Scientist and a Group Leader with the Agency for Science, Technology and Research (A$^\star$STAR), Singapore. His research interests include the automatic and para-linguistic analysis of speaker characteristics, ranging from speaker recognition, language and accent recognition, voice biometrics, spoofing, and countermeasures. He was the recipient of the Singapore IES Prestigious Engineering Achievement Award 2013 and the Outstanding Service Award by IEEE ICME 2020. Since 2016, he has been an Editorial Board Member of Elsevier Computer Speech and Language. From 2017 to 2021, he was an Associate Editor for IEEE/ACM Transactions on Audio, Speech, and Language Processing. He is an elected Member of the IEEE Speech and Language Processing Technical Committee and was the General Chair of the Speaker Odyssey 2020 Workshop.
\end{IEEEbiography}

\vspace{-20 pt}

\begin{IEEEbiography}[{\includegraphics[width=1in,height=1.25in,clip,keepaspectratio]{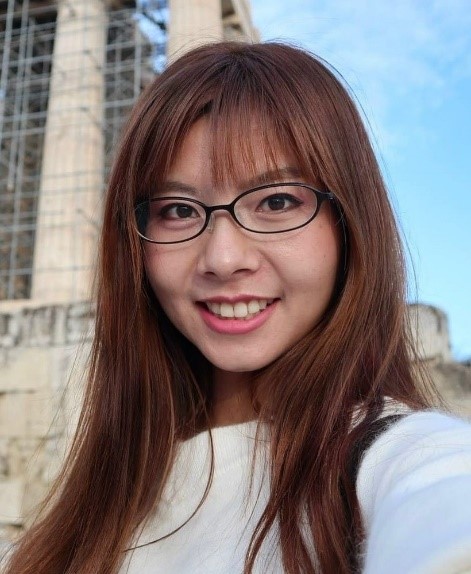}}]{Qiongqiong Wang} (Member, IEEE)  is currently a Lead Research Engineer at the Institute for Infocomm Research (I$^2$R), Agency for Science, Technology and Research (A$^\star$STAR), Singapore.  She was a researcher at the Biometrics Research Laboratories, NEC Corporation, Japan, from 2013 to 2021. She received her M.E in computer science from the Tokyo Institute of Technology, Japan, in 2013, and B.E from the Undergraduate School of Physics, Shanghai Jiao Tong University, China, in 2011. Her research focuses on speaker recognition, deception detection, emotion recognition, speech anti-spoofing, speech enhancement, large language model (LLM), and speech foundation model. 
\end{IEEEbiography}

\vspace{-20 pt}

\begin{IEEEbiography}[{\includegraphics[width=1in,height=1.25in,clip,keepaspectratio]{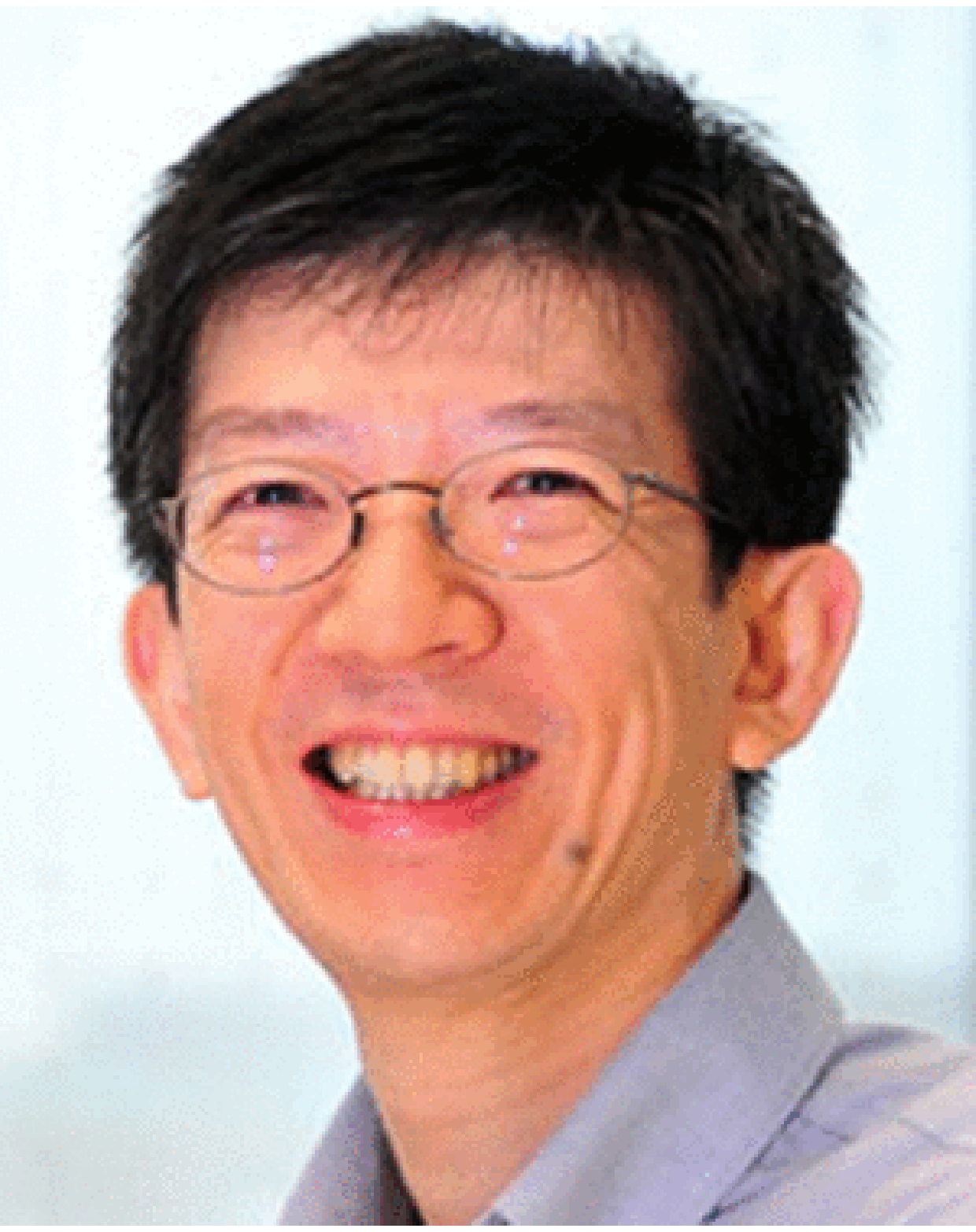}}]{Haizhou Li} (Fellow, IEEE) received the B.Sc., M.Sc., and Ph.D. degrees in electrical and electronic engineering from the South China University of Technology, Guangzhou, China, in 1984, 1987, and 1990 respectively. He is currently a Presidential Chair Professor and the Executive Dean of the School of Data Science, The Chinese University of Hong Kong, Shenzhen, China. He is also an Adjunct Professor with the Department of Electrical and Computer Engineering, National University of Singapore, Singapore. Prior to that, he taught with The University of Hong Kong, Hong Kong, during 1988–1990, and South China University of Technology, during 1990–1994. He was a Visiting Professor with CRIN, France, during 1994–1995, Research Manager with the AppleISS Research Centre during 1996–1998, the Research Director with Lernout \& Hauspie Asia Pacific during 1999–2001, Vice President with InfoTalk Corporation Ltd. during 2001–2003, and Principal Scientist and Department Head of human language technology with the Institute for Infocomm Research, Singapore, during 2003–2016. His research interests include automatic speech recognition, speaker and language recognition, natural language processing. Dr. Li was the Editor-in-Chief of IEEE/ACM TRANSACTIONS ON AUDIO, SPEECH AND LANGUAGE PROCESSING during 2015–2018, an elected Member of IEEE Speech and Language Processing Technical Committee during 2013–2015, the President of the International Speech Communication Association during 2015–2017, President of Asia Pacific Signal and Information Processing Association during 2015–2016, and President of Asian Federation of Natural Language Processing during 2017–2018. Since 2012, he has been a Member of the Editorial Board of \textit{Computer Speech and Language}. He was the General Chair of ACL 2012, INTERSPEECH 2014, ASRU 2019 and ICASSP 2022. Dr. Li is a Fellow of the ISCA, and a Fellow of the Academy of Engineering Singapore. He was the recipient of the National Infocomm Award 2002, and President’s Technology Award 2013 in Singapore. He was named one of the two Nokia Visiting Professors in 2009 by the Nokia Foundation, and U Bremen Excellence Chair Professor in 2019.
\end{IEEEbiography}



\vfill

\end{document}